\documentclass[%
reprint,
superscriptaddress,
 amsmath,amssymb,
 aps,
prb,
]{revtex4-2} 
\usepackage{chemformula} 
\usepackage[T1]{fontenc} 
\usepackage{graphicx}
\usepackage{bm}
\usepackage[utf8]{inputenc}
\usepackage[T1]{fontenc}
\usepackage{mathptmx}
\usepackage{etoolbox}
\usepackage{braket}
\usepackage{upgreek}
\usepackage{multirow}

\begin{document}
\title{Exploring Overlapping Mechanisms of Dynamic Nuclear Polarization in Type 1b HPHT Diamond}
\author{Brendan C.~Sheehan}
\email[Corresponding author: ]{brendan.c.sheehan@dartmouth.edu}
\affiliation{Department of Physics and Astronomy, Dartmouth College, Hanover, NH 03755, USA}
\author{Margaret Hubble}
\altaffiliation[Current address: ]{Department of Physics, CUNY-City
College of New York, New York, NY 10031, United
States}
\affiliation{Department of Physics and Astronomy, Dartmouth College, Hanover, NH 03755, USA}
\author{Daphna Shimon}
\affiliation{Institute of Chemistry, The Hebrew
University of Jerusalem, Jerusalem 9190401, Israel}
\author{Chandrasekhar Ramanathan}
\email[Corresponding author: ]{chandrasekhar.ramanathan@dartmouth.edu}
\affiliation{Department of Physics and Astronomy, Dartmouth College, Hanover, NH 03755, USA}

\date{\today}

\begin{abstract}
The inhomogeneous distribution of P1 centers in type 1b HPHT diamond samples allows multiple DNP mechanisms to occur within the same crystal, resulting in complex DNP spectra.  At some crystal orientations, different DNP mechanisms can compete to drive hyperpolarization with different signs at the same applied microwave frequency. We perform microwave-irradiated DNP using both monochromatic and frequency-modulated microwave excitation to explore the competition between these DNP mechanisms in diamond at room temperature.  We demonstrate that frequency-modulated DNP is a tool for suppressing certain DNP mechanisms while enhancing others in a single-crystal diamond sample. Frequency modulation also enables higher enhancement of the NMR signal beyond traditional monochromatic DNP under some conditions. In a powder sample, competing enhancement mechanisms can also arise from different crystallite orientations in the powder.  We observe that at certain microwave frequencies the DNP signal changes sign during the polarization build-up, even with monochromatic microwave irradiation. We do not observe this phenomenon in any single-crystal spectrum. We discuss both methods of investigating competing mechanisms of DNP as a means of selectively enhancing different DNP mechanisms driving $^{13}$C NMR signal enhancement.
\end{abstract}
\maketitle
\newpage

\section{Introduction}

Paramagnetic defects in diamond are a class of defects offering strong DNP-enhanced NMR to nearby $^{13}$C spins. The negatively charged nitrogen-vacancy (NV) center in diamond, for example, has been shown to provide large DNP enhancement at both cryogenic and room temperatures after optical polarization~\cite{kingOpticalPolarization13C2010,kingRoomtemperatureSituNuclear2015}. While this is typically performed by pumping at low magnetic field and rapidly shuttling the sample into high magnetic field (> 1 T) for NMR readout~\cite{ajoyEnhancedDynamicNuclear2018,ajoyHyperpolarizedRelaxometryBased2019,ajoyLowfieldMicrowavemediatedOptical2021}, large signal enhancements have also been observed at high magnetic fields \cite{scottPhenomenologyOpticallyPumped2016}. Though less frequently studied, the P1 center in diamond, a substitutional nitrogen lattice defect with long spin-lattice relaxation and coherence times at room temperature~\cite{reynhardtDynamicNuclearPolarization1998,cox13C14N15N1994,takahashiQuenchingSpinDecoherence2008}, can provide an analogously useful source of room-temperature DNP enhancement without the need for optical pumping. P1 centers have been shown to generate large DNP enhancement both at cryogenic~\cite{casabiancaFactorsAffectingDNP2011,vonwitteTemperatureDependentDynamicNuclear2025} and room temperatures~\cite{shimonLargeRoomTemperature2022,shimonRoomTemperatureDNP2022}, and provide insight into the dynamics of polarization transfer that occur at variable electron-spin clustering~\cite{bussandriP1CenterElectron2024,nir-aradPeculiarEPRSpectra2024,nir-aradNitrogenSubstitutionsAggregation2024,nir-aradClusteringHeterogeneousP12025,sternP1CenterNetwork2025}.

\begin{figure*}[ht!]
    \centering
   \includegraphics[width=0.95\linewidth]{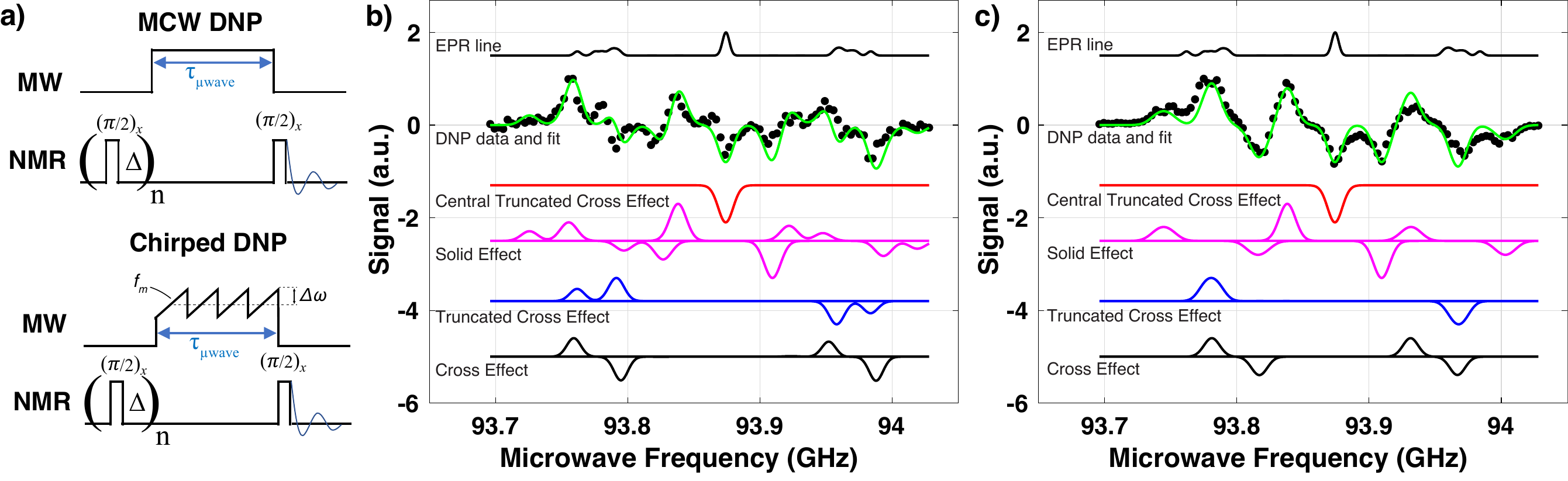}
\caption{\textbf{DNP spectrum and spectral decomposition for two orientations of a single-crystal diamond sample under monochromatic MW irradiation.} (a) Pulse sequences used in this work for monochromatic MW irradiation (top) and frequency-modulated (chirped) MW irradiation (bottom). The chirp can be described by the central frequency of the chirp ($\omega_{MW}$), the modulation amplitude $\Delta\omega$, and the modulation frequency (slope) $f_m$. For all experiments a saturation train of 30 $\uppi/2$ pulses was used, with a saturation delay between each pulse in the train of $\Delta = 50~\upmu$s. (b) Orientation 1 shows a spectrum containing multiple frequencies with competition between the positive and negative enhancements provided by different DNP mechanisms. This spectrum has a MW buildup time of 100~s. (c) Orientation 2 shows multiple DNP mechanisms summing constructively to generate large DNP enhancement, though few MW frequencies include competition in the sign of the signal. Here, the MW buildup time was 200~s. The top line on each panel is the EPR line, simulated with EasySpin~\cite{stollEasySpinComprehensiveSoftware2006}, with the largest intensity peak normalized to one.}
    \label{fig:crystaldecomp}
\end{figure*}

The P1 center in diamond can be described to a good approximation with the single-spin Hamiltonian
\begin{equation}
\label{eq:spinham}
    \mathcal{H} = \mu_B\hat{S}\cdot\mathbf{g}\cdot\mathbf{B} + \hat{I}\cdot\mathbf{A}\cdot\hat{S},
\end{equation}
with a Zeeman interaction characterized by $\mathbf{g} = [2.0023,2.0023,2.00225]$. The anisotropic hyperfine coupling between the spin-1 nitrogen and electron is described by $\mathbf{A} = [81.3,81.3,114]$~MHz. Below 140~GHz the $g$-anisotropy of the P1 center is not resolvable via EPR. DNP using the P1 center as a polarization reservoir typically generates an NMR signal enhancement above the thermal $^{13}$C NMR signal on the order of 100 at low microwave power and 3.34~T~\cite{shimonLargeRoomTemperature2022}; at 7~T enhancements as high as 1500 have been reported with 0.5~W microwave power~\cite{nevzorovMultiresonantPhotonicBandgap2018}. Because of the anisotropic hyperfine line and the tendency of P1 centers in diamond to form both isolated and clustered pockets within a single crystal~\cite{bussandriP1CenterElectron2024}, the DNP spectrum, the DNP enhancement measured as a function of microwave (MW) frequency, features substantial complexity and has been shown to be made up of numerous polarization transfer mechanisms~\cite{shimonLargeRoomTemperature2022,nir-aradNitrogenSubstitutionsAggregation2024,palaniDynamicNuclearPolarization2024,nir-aradClusteringHeterogeneousP12025,sternP1CenterNetwork2025}.

While DNP spectra observed using P1 centers in diamond typically correlate reasonably well with their EPR lines, the full DNP spectrum represents a sum of multiple different physical effects. Each component mechanism has a characteristic rate of DNP buildup, as well as a MW frequency behavior related to the EPR line. The anisotropy of the hyperfine interaction generates strongly orientation-dependent satellites in the EPR line. The condition under which the EPR signal is obtained can play an important role in determining which population of P1 spins is measured~\cite{equbalBalancingDipolarExchange2020}.

The DNP spectrum in diamond can be decomposed into the sum of different components arising from the solid effect (SE), the cross effect (CE), and the truncated cross effect (tCE). The SE is typically an isolated electron-nuclear process wherein anisotropic hyperfine interactions mix nuclear spin states. At higher electron spin concentrations, the CE arises from three-spin systems, particularly when the EPR line is inhomogeneously broadened. If one of the two populations of electrons has a significantly shorter $T_1$ relaxation time, DNP is often observable only when irradiating the population with slower $T_1$ relaxation, truncating three-spin the polarization transfer efficiency -- the tCE~\cite{equbalTruncatedCrossEffect2018}. In our initial work, we also noted the presence of an "apparent Overhauser effect" which occurred at the central electron Larmor frequency~\cite{shimonLargeRoomTemperature2022}. More recent work has shown that this is actually a tCE between isolated and strongly exchange-coupled clusters of P1 centers in these HPHT diamonds~\cite{bussandriP1CenterElectron2024}.

In this work, we demonstrate control over the mechanism of P1 center DNP enhancement of $^{13}$C NMR signal in diamond at MW frequencies exhibiting competing mechanisms of electron polarization transfer at low microwave power. The experiments are performed on both powder and single-crystal HPHT diamond at room temperature. In Section~\ref{sec:freqmod} we extend our previous work on chirped DNP of HPHT diamond powder to a single-crystal sample, presenting results of frequency-modulated (chirped) DNP on the single-crystal sample at 3.34 T~\cite{shimonRoomTemperatureDNP2022}. The schematic of the DNP experiments are shown in Figure~\ref{fig:crystaldecomp}(a). The orientation-dependent nature of the EPR and DNP spectra of the P1 center in a single crystal provides better spectral resolution of individual peaks than in powder~\cite{shimonLargeRoomTemperature2022}. We harness these sharp features to improve DNP signal with frequency modulation. Controlling the crystal orientation allows us to identify conditions under which DNP mechanisms overlap. We use periodic linear microwave frequency chirping with differing modulation amplitudes ($\Delta\omega$) and modulation frequencies ($f_m$) to interrogate the physics of when frequency modulated microwaves most efficiently enhance DNP signal~\cite{kisselevModulationEffectDynamic1995,thurberLowtemperatureDynamicNuclear2010,bornetMicrowaveFrequencyModulation2014}. We compare these to previous results on the powder sample and use the two samples to understand the physics governing polarization transfer from clustered P1 centers in powder and single-crystal HPHT diamond~\cite{shimonRoomTemperatureDNP2022}. We note that in all DNP experiments in this manuscript we are power-limited; we observe that the DNP dependence on MW power is linear for all mechanisms at 240~mW (the nominal power used in this work). Power saturation in the DNP enhancement is observed for all mechanisms except the SE at 500~mW~\cite{shimonLargeRoomTemperature2022}.

In Section~\ref{sec:signalinversion} we characterize the buildup time dependence of the DNP spectrum in powdered HPHT diamond, under monochromatic continuous wave (MCW) MW irradiation at both 3.34~T and 7.05~T. At certain MW frequencies we observe positive DNP signal enhancement for short buildup time, but inverted DNP signal enhancement at long buildup time. Through spectral decomposition we attribute this effect to competition between the CE and SE of the low-frequency hyperfine manifold hosted in different crystallites in the powder. We do not observe this effect in single-crystal HPHT diamond.

\begin{figure*}[ht!]
    \centering
   \includegraphics[width=0.95\linewidth]{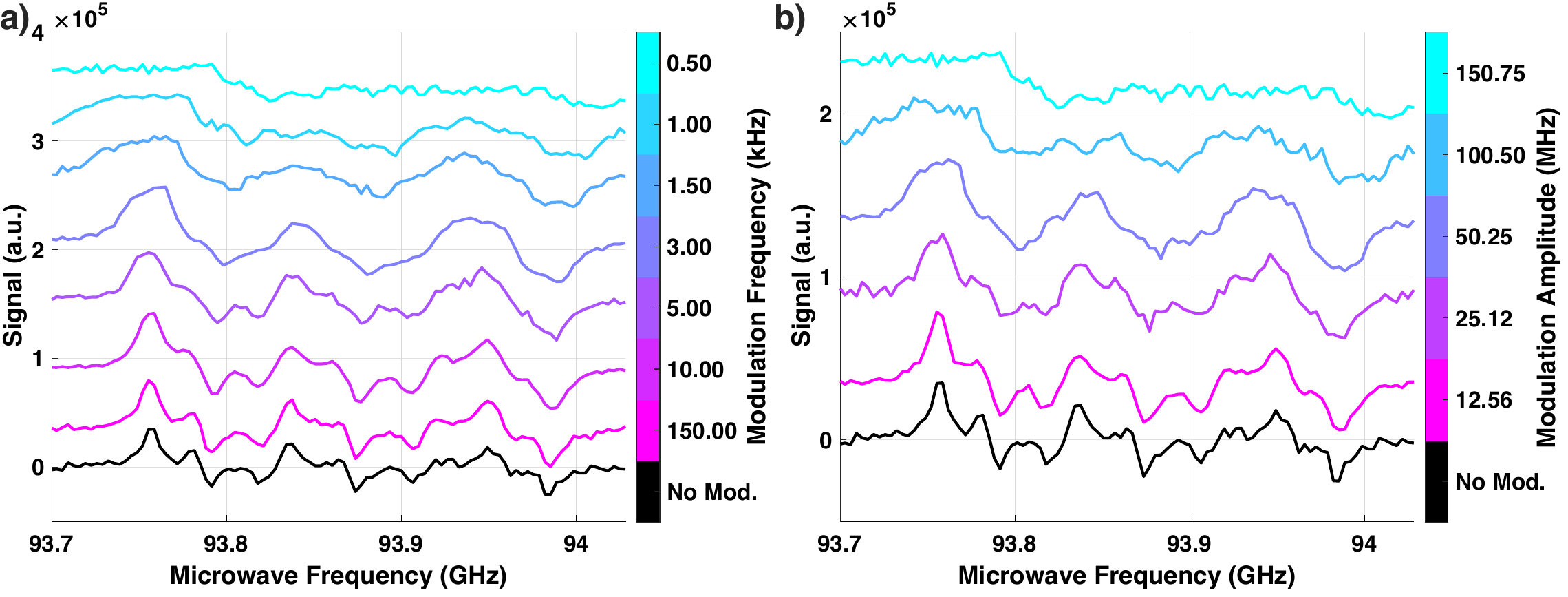}
\caption{\textbf{DNP spectra using frequency- and amplitude-modulated microwaves, measured on Orientation 1.} (a) A waterfall plot of the DNP spectrum using modulated microwaves. Each trace was measured using a modulation amplitude of $\Delta\omega = 150.75$~MHz and a different value of $f_m$ (colorbar). (b) A waterfall plot at constant $f_m$ (500~Hz) and varying $\Delta\omega$ (colorbar). In both panels each trace was measured using a MW buildup time of 100~s.}
    \label{fig:modwater}
\end{figure*}

\section{Microwave frequency modulation of a single-crystal diamond P1-DNP spectrum}
\label{sec:freqmod}

We use two orientations of a single crystal sample to examine the physics of modulated polarization transfer under different combinations of DNP transfer mechanisms, and compare these data to our previous study of powdered diamond under frequency-modulated DNP~\cite{shimonRoomTemperatureDNP2022}. Figure~\ref{fig:crystaldecomp} shows a comparison of the DNP spectrum to the EPR lines and the decomposition of the different DNP mechanisms can be seen for a crystal sample at the two orientations. Orientation 1 contains many frequencies with competing mechanisms of DNP enhancement. Orientation 2, in contrast, features a DNP spectrum showing only very small overlap regions where individual mechanisms in the spectral decomposition have opposite sign. In each panel we employ a fitting technique to ensure proper decomposition; the fitting methodology is described in Section~\ref{sec:meth} and in our previous publication~\cite{shimonLargeRoomTemperature2022}.

Our previous work has investigated the control of DNP mechanisms through frequency modulation in a powder sample~\cite{shimonRoomTemperatureDNP2022}. Here, we examine the effects of narrow-band to broadband frequency modulation at several central MW frequencies for both orientations shown in Figure~\ref{fig:crystaldecomp}. Figure~\ref{fig:modwater} shows modulated DNP spectra at different values of modulation amplitudes $\Delta\omega$ and frequencies $f_m$, for Orientation 1. Panel (a) shows a set of spectra at different values of modulation frequency (left side of panel) from 0.5~kHz to 150~kHz, measured at a modulation amplitude of 150.75~MHz. Panel (b) shows DNP spectra where the modulation amplitude is increased from 12.56~MHz to 150.75~MHz, at a fixed modulation frequency of 0.5~kHz. In both cases the MCW spectrum is shown at the bottom of the panel. 

In the case of maximal $\Delta\omega$, as $f_m$ is varied, the DNP spectrum resembles the nonmodulated case at large $f_m$, but begins to become distorted as $f_m$ decreases -- an observation consistent with our previous work studying modulated DNP spectra in powdered diamond~\cite{shimonRoomTemperatureDNP2022}. The threshold for substantial change in the spectrum coincides with $f_m\sim \mathcal{O}(T_{1e})$. At modulation rates faster than $T_{1e}$, the modulated spectra begin to more closely resemble the nonmodulated spectrum due to the fast passage of the microwave chirp behaves like a CW tone at each MW frequency. As $f_m$ slows compared to $T_{1e}$, however, the electron population is able to track the modulation, transferring polarization and relaxing during the chirp before becoming re-polarized during the following chirp. Polarization can therefore be transferred across the entire bandwidth of the chirp, and the resulting DNP signal is the summed contribution from all mechanisms driving polarization transfer within the bandwidth.

We previously characterized biexponential $T_{1e}$ decay at room temperature and 2.5~GHz for the powder (110~$\upmu$s and 1.3~ms) and the single-crystal (70~$\upmu$s and 1.3~ms) diamond samples~\cite{shimonLargeRoomTemperature2022}; we denote these $T_{1e,\text{fast}}$ and $T_{1e,\text{slow}}$. Though this measurement was performed at low field, we note that P1 spin-lattice relaxation at room temperature remains in the same order-of-magnitude up to W-band~\cite{terblancheRoomtemperatureFieldDependence2000} and has consistently been measured to be of order 1~ms in type 1b diamonds across a range of P1 center concentrations similar to the sample examined in this study~\cite{reynhardtTemperatureDependenceSpinspin1998}.

At very slow $f_m$, where adiabatic passage allows more efficient polarization transfer, the modulated spectrum tends toward an integration of the contribution to the DNP signal within the band of the chirp. This effect is most clearly seen in the case shown in Figure~\ref{fig:modwater}(b), where increasing modulation amplitude has the effect of sampling resonances across the entire EPR line. The effect is most noticeable when the width of the chirp $2\Delta\omega\gtrsim\omega_C$, where $\omega_C$ is the nuclear Larmor frequency. 

\begin{figure*}[ht!]
    \centering
   \includegraphics[width=1\linewidth]{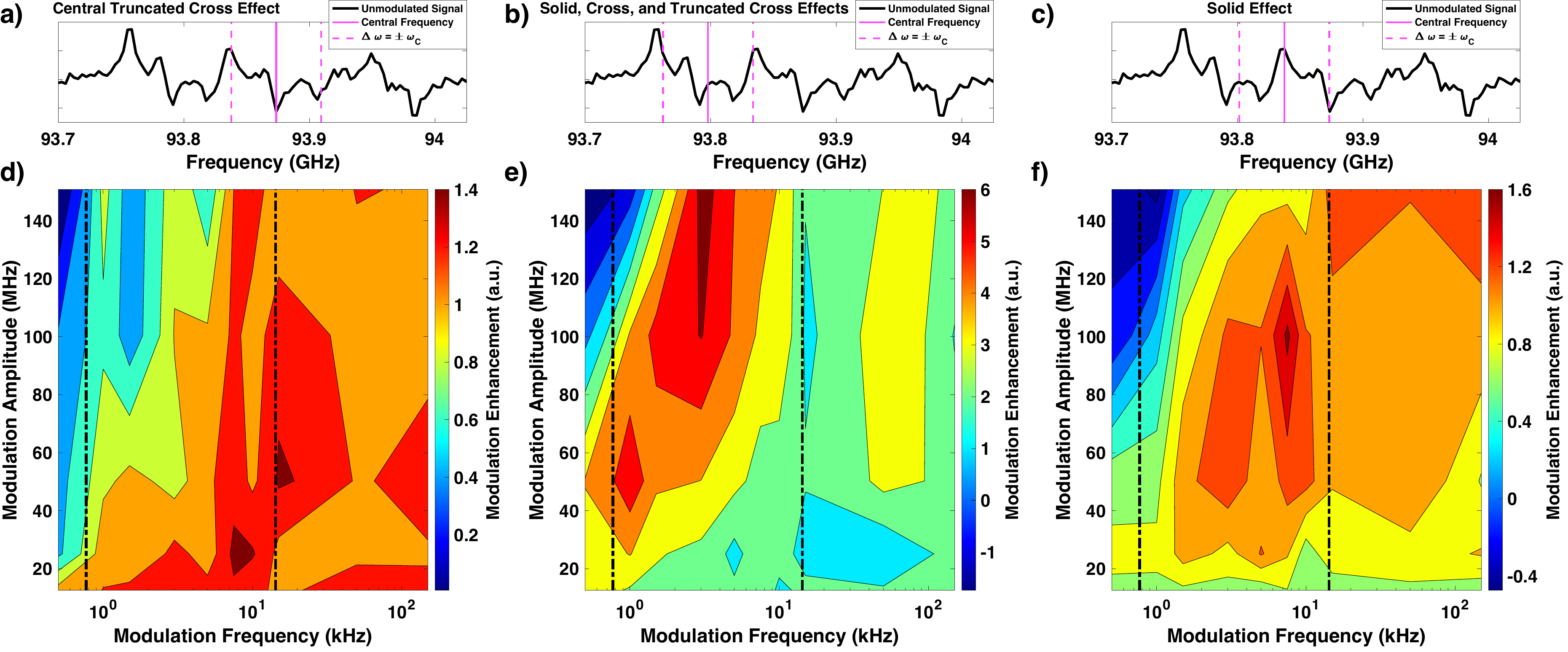}
\caption{\textbf{Colormaps of DNP signal enhancement for varying modulation frequency and amplitude for Orientation 1.} Panels (a), (b), and (c) reproduce the DNP spectrum from Figure~\ref{fig:crystaldecomp}(b) with central MW irradiation frequency shown in solid pink. The dashed pink lines represent the edges of MW irradiation when the modulation amplitude $\Delta\omega$ equals the $^{13}$C Larmor frequency $\omega_C$ (35.8~MHz). Panels (d), (e), and (f) show the corresponding colormap of modulation frequency and amplitude for the central frequency shown in (a), (b), and (c), respectively. In each colormap the two-timescale $1/T_{1e}$ relaxation rates, measured at 2.5~GHz, are drawn as vertical dashed black lines.}
    \label{fig:ModMarge}
\end{figure*}

\begin{figure*}[ht!]
    \centering
   \includegraphics[width=1\linewidth]{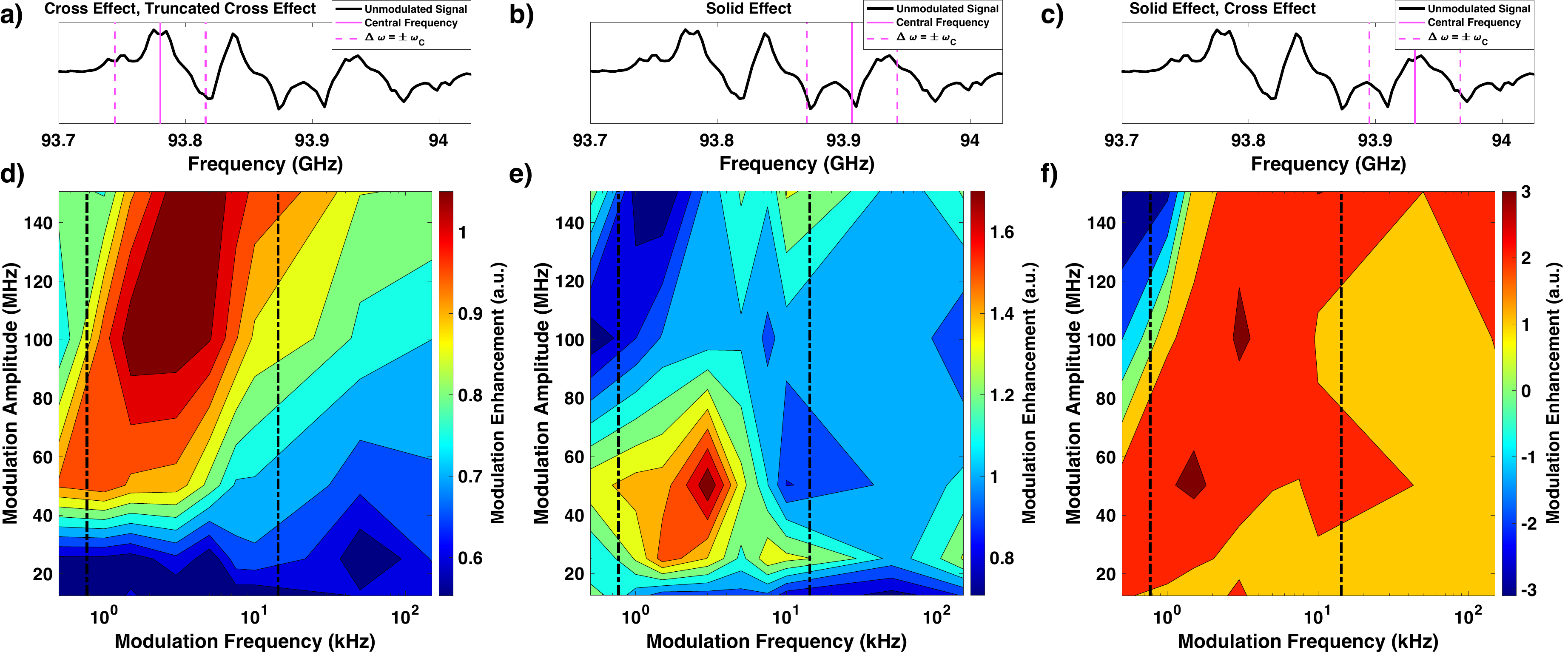}
\caption{\textbf{Colormaps of DNP signal enhancement for varying modulation frequency and amplitude for Orientation 2.} Panels (a), (b), and (c) reproduce the DNP spectrum from Figure~\ref{fig:crystaldecomp}(a) with central MW irradiation frequency shown in solid pink. The dashed pink lines represent the edges of MW irradiation when the modulation amplitude $\Delta\omega$ equals the $^{13}$C Larmor frequency $\omega_C$ (35.8~MHz). Panels (d), (e), and (f) show the corresponding colormap of modulation frequency and amplitude for the central frequency shown in (a), (b), and (c), respectively. In each colormap the two-timescale $1/T_{1e}$ relaxation rates, measured at 2.5~GHz, are drawn as vertical dashed black lines.}
    \label{fig:ModBrendan}
\end{figure*}

Figures~\ref{fig:ModMarge} and~\ref{fig:ModBrendan} each show three examples of colormaps of the relative enhancement due to frequency modulation at a fixed central microwave frequency, where the DNP signal was measured over a grid of modulation amplitudes $\Delta\omega$ and frequencies $f_m$. The modulation frequency $f_m$ sampling was performed using ten values with roughly logarithmic spacing from 0.5~kHz to a maximum of 150~kHz, while $\Delta\omega$ was sampled with five values from 12.56~MHz to 150.75~MHz. Figure~\ref{fig:ModMarge} references Orientation 1, while Figure~\ref{fig:ModBrendan} references Orientation 2. 

\subsection{Frequency-modulated DNP at orientation 1}

Orientation 1 (Fig.~\ref{fig:crystaldecomp}(b)) contains SE, CE, and tCE contributions to signal enhancement. Figures~\ref{fig:ModMarge}(a),~(b), and~(c) reproduce the MCW DNP spectrum. Figures~\ref{fig:ModMarge}(d),~(e), and~(f) show colormaps of $\Delta\omega$ and frequencies $f_m$ at the central frequency marked with the solid pink vertical line in the panel above. Dashed pink vertical lines in the top row mark the frequency range corresponding to a modulation amplitude of $\Delta\omega=\omega_C$, the $^{13}$C Larmor frequency, on either side of the central frequency (for a full irradiation bandwidth of $2\omega_C$). Frequency-modulated signal is scaled to the MCW DNP signal recorded at the central frequency; no additional enhancement due to frequency modulation yields a modulation enhancement of 1. 

Figures~\ref{fig:ModMarge}(a) and~(d) show the result of probing frequency-modulated DNP at a central frequency matching the central tCE (and the $m_I=0$ EPR manifold). In the colormap Figure~\ref{fig:ModMarge}(d), modulation enhancement reaches a maximum of 1.4, with intermediate modulation amplitude and a modulation rate faster than $T_{1e,\text{fast}}$. At modulation amplitudes matching $\Delta\omega\approx\omega_C$, the chirp samples both the zero-quantum and double-quantum SE peak; this broad sampling of the SE line is an example of integrated solid effect (iSE) frequency modulation~\cite{henstraEnhancedDynamicNuclear1988,henstraDynamicNuclearPolarisation2014,canFrequencySweptIntegratedStretched2018}. We observe a maximal modulation enhancement when the MW frequency is parked at the central EPR manifold and the chirp samples both SE peaks, suggestive of iSE-enhanced DNP signal at $\Delta\omega\approx\omega_C$ and intermediate $f_m$.

Figures~\ref{fig:ModMarge}(b) and~(d) show a colormap of frequency-modulated DNP at a central frequency where the SE and CE provide a negative DNP signal, but the tCE provides a positive signal -- an example of a MW frequency where competing mechanisms of DNP enhancement exist. We observe not only a maximum modulation enhancement of a factor of six over the MCW DNP signal, but also see an inversion of the DNP signal to a modulation enhancement of -1.5 at large, slow modulation. We note that while the modulation enhancement is large, the MCW DNP signal is actually fairly small at this central frequency $\omega_{MW}$, as compared to the signals probed in Figures~\ref{fig:ModMarge}(d) and~(f). Figures~\ref{fig:ModMarge}(c) and~(f) show the results of probing at the positive (zero-quantum) SE peak of the $m_I=0$ central EPR line. We again observe an inversion of the DNP signal at large, slow modulation.

\subsection{Frequency-modulated DNP at orientation 2}

Orientation 2 (Fig.~\ref{fig:crystaldecomp}(b)) contains SE, CE, and tCE contributions to signal enhancement, though in a different configuration than Orientation 1. Figures~\ref{fig:ModBrendan}(a) and~(d) show the result of probing frequency-modulated DNP at a frequency where the $m_I = 1$ hyperfine line provides positive enhancement from both the CE and tCE. Here we see inefficient polarization transfer upon application of modulation at nearly all values of $\Delta\omega$ and $f_m$, excepting large, intermediate-rate modulation. The modulation enhancement is maximized at slightly greater than $\pm1$.

Figures~\ref{fig:ModBrendan}(b) and~(e) show the result of probing frequency-modulated DNP at the negative (double-quantum) SE peak in the line. Though moderately strong modulation ($\Delta\omega\approx 50$~MHz) can achieve a $\sim$70\% modulation enhancement of the DNP signal, a slow-but-strong modulation actually achieves a decrease in signal, likely because the bandwidth in this case samples such a large portion of the spectrum as to integrate the DNP signal to near zero.

Figures~\ref{fig:ModBrendan}(c) and~(f) show the result of probing frequency-modulated DNP at a frequency where the $m_I = -1$ hyperfine line provides positive enhancement from both the SE and CE. By extending $\Delta\omega$ broad enough, we observe a modulation enhancement of nearly a factor of three. Increasing $\Delta\omega$ beyond $\sim80$~MHz, however, while modulating slowly, actually inverts the DNP enhancement, to a maximum negative enhancement of -3. The modulation enhancement range in this experiment, from -3 at large slow modulation, to 3 at $\Delta\omega\approx100$~MHz and $f_m\approx2$~kHz, is an example of the frequency modulation technique's ability to choose which DNP mechanisms contribute to the NMR signal. Setting $\Delta\omega\approx145$~MHz includes nearly the entire DNP line, excepting the lowest-frequency portion of the $m_I=-1$ manifold (which typically generates positive enhancement). At this modulation amplitude and a slow enough modulation rate ($1/f_m$), the modulation rate is slower than $T_{1e,\text{slow}}$, meaning the effective MW power is too low to saturate the DNP transitions at $\omega_{MW}$~\cite{kunduDNPMechanisms2019}. Rather, the chirp integrates over the DNP spectrum and produces an inverted DNP signal.

\subsection{Solid effect and cross effect under MW frequency modulation}

Figures~\ref{fig:ModMarge}(b),~\ref{fig:ModMarge}(c),~\ref{fig:ModBrendan}(b), and~\ref{fig:ModBrendan}(c) show colormaps with the central MW frequency parked at a frequency corresponding (at least in part) to solid effect DNP. In all four of these datasets, the two $T_{1e}$ relaxation times (70~$\upmu$s and 1.3~ms, measured in~\cite{shimonLargeRoomTemperature2022}) set characteristic timescales for frequency-modulated dynamics. As a guide to the eye, each colormap in Figures~\ref{fig:ModBrendan} and~\ref{fig:ModMarge} has vertical dashed lines demarking the two relaxation rates $1/T_{1e}$. As has previously been discussed, use of a modulation rate slower than $T_{1e}$ allows the bath to be reset faster than the chirp; this effect becomes particularly dramatic at modulation rates slower than the longer timescale. We suggest the falloff of DNP signal beyond this threshold time may be driven by isolated P1 centers~\cite{shimonRoomTemperatureDNP2022}. The shorter $T_{1e}$ timescale likely represents either electron polarization reorganization driven by electron-electron spectral diffusion (eSD)~\cite{leavesleyEffectElectronSpectral2017,shimonTransitionSolidEffect2022} or the distinctly faster recovery time of clustered P1 centers, present in the DNP dynamics of single-crystal HPHT diamond~\cite{nir-aradClusteringHeterogeneousP12025}.

The contributions to the DNP signal from the two-spin (electron-nuclear) SE and three-spin (electron-electron-nuclear) CE can be compared to large spin system simulations performed in our previous work~\cite{shimonRoomTemperatureDNP2022}. There, we simulated the DNP enhancement for the SE and CE as a function of the $\Delta\omega$ for three cases: negligible (slow) eSD, intermediate eSD, and large (fast) eSD. Comparing the poor DNP enhancement from frequency-modulated SE and CE data in the powdered diamond sample to the simulations, we found that the powder sample likely contains intermediate or even large eSD. For large eSD, modulation does not improve the DNP contribution of a single electron to either the SE or CE signal; the DNP enhancement simply decreases. 

For negligible eSD, the amplitude $\Delta\omega$ must be studied. At small $\Delta\omega<\omega_C$, selection of a central MW frequency $\omega_{MW}$ lets us irradiate the DQ and ZQ transitions that obey $\omega_{MW}\pm\omega_C$, while partially saturating SQ electrons at $\omega_{MW}$. When $\Delta\omega\approx\omega_C$, the SQ and DQ/ZQ transitions of the same electron begin to saturate; this causes the DNP enhancement to drop sharply above $\Delta\omega\approx2\omega_C$. We therefore expect DNP enhancement to be maximized when $\Delta\omega\approx2\omega_C$ (or slightly lower than $\omega_C$) if eSD is negligible. Intermediate eSD creates a scenario where the modulation enhancement of the DNP signal occurs at a lower $\Delta\omega$ than $2\omega_C$.

\begin{figure*}[ht!]
    \centering
   \includegraphics[width=0.95\linewidth]{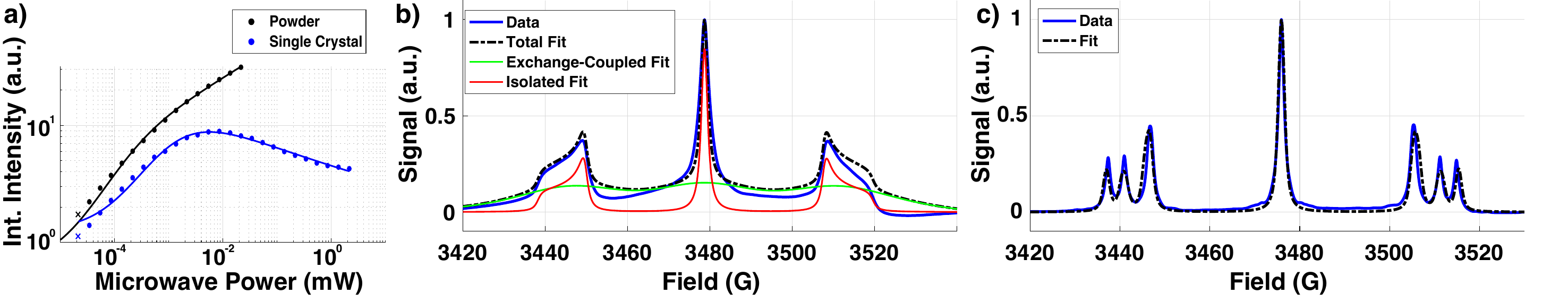}
\caption{\textbf{X-band EPR spectra of powder and single-crystal diamond at room temperature.} (a) Twice-integrated EPR signal as a function of microwave power for the powder sample (black) and single-crystal (blue). Lines are fits of the data to a saturation curve. The spectra measured at lowest power, shown with an "x", for each sample (0.034~$\upmu$W) are shown in panels (b) (powder) and (c) (single-crystal). (b) Integrated EPR spectrum of the powder sample at X-band shows a typical powder pattern of the P1 center albeit with a broad background visible upon integration. (c) Integrated EPR spectrum of the single-crystal diamond P1 center shows clear hyperfine satellites, but without the broad background feature. Simulations were performed with EasySpin~\cite{stollEasySpinComprehensiveSoftware2006}. In (b) and (c) the largest intensity peak was normalized to one.}
    \label{fig:xband}
\end{figure*}

In this single-crystal sample, we observe maximal modulation enhancement when $\omega_{MW}$ is centered on a $m_I=0$ SE peak when $\Delta\omega$ is slightly smaller than $2\omega_C = 72$~MHz (Fig.~\ref{fig:ModBrendan}(b)) or slightly larger than $2\omega_C$ (Fig.~\ref{fig:ModMarge}(c)). In both cases, since $\Delta\omega\sim2\omega_C$, we suggest that this single-crystal diamond sample actually has quite small (slow) electron spectral diffusion. We find this eSD to be much slower (weaker) than in the powdered HPHT sample studied in our previous work.

\subsection{EPR lines show exchange coupling and provide an indirect measure of electron spectral diffusion}

We turn to CW EPR as a means of characterizing the EPR linewidths and exchange coupling observed in each sample. The linewidths provide an indirect measure of eSD; the strength of the exchange coupling provides a mechanism by which eSD is mediated. We show the power saturation curves and EPR lines of the powder and single-crystal spectra in Figure~\ref{fig:xband}, acquired using standard field-modulated EPR at X-band. The power saturation curves (a) ensure both samples (powder, black; single-crystal, blue) begin to saturate above a CW power of $10^{-4}$~$\upmu$W. The solid lines are power saturation curves fit to the twice-integrated spectra (dots). The "x" at the lowest powers for each sample are the spectra shown in panels (b) and (c).

Figure~\ref{fig:xband}(b) shows the integrated (zeroth harmonic) EPR line of the powder sample in blue. Integration reveals a broad background which does not disappear even in the low-power regime. We ascribe this background to P1 clusters which have strong exchange-coupling. Bussandri and coworkers first made this observation at high field (7~T) using pulsed EPR~\cite{bussandriP1CenterElectron2024}. Here we extend the EPR characterization of the exchange-coupled population in HPHT diamond powder to lower field.

We fit the powder spectrum using a custom two-component fit based on EasySpin; the isolated component uses the Hamiltonian Eq.~\ref{eq:spinham} and is shown in red, while the exchange-coupled component explains the broad background (green). Fitting parameters are homogeneous and inhomogeneous broadening for each population of spins, exchange coupling (for one population), and a scaling term. The strength of the exchange coupling is found to be 24(6)~MHz. Results of the fit are summarized in the first row of Table~\ref{table:eprfits}. These results produce a simulated spectrum in good qualitative agreement with the high-field results~\cite{bussandriP1CenterElectron2024}.

The single-crystal spectrum and best fit are shown in Figure~\ref{fig:xband}(c). Notably, no broad background is observed -- suggestive of much smaller exchange coupling (or perhaps few to no exchange-coupled clusters) in the sample. The best-fit line (black) is the result of fitting to Eq.~\ref{eq:spinham}. We note that the relative amplitudes of the hyperfine satellites at this orientation are only properly explained by a model containing a twinned crystal. Twinning is not visible in the EPR spectrum at all orientations; we do not observe evidence of twinning in this sample in our DNP spectra. Addition of an exchange-coupled term into the Hamiltonian shows only modest improvement in the fit; we find the strength of the coupling to be 1.3(0)~MHz. Fitting parameters are homogeneous and inhomogeneous broadening, exchange coupling strength, Euler angles to represent the orientation of the sample relative to the field, and a scaling of the relative weight of the twinned populations. Results of the fit are summarized in the second row of Table~\ref{table:eprfits}.

\begin{table}[ht!] 
\centering 
\begin{tabular}{|c|c|c|} 
\hline 
 & \textbf{Fitting Parameters} & \textbf{Best-Fit Values} \\ 
\hline
\hline 
\textbf{Powder} & \begin{tabular}[t]{@{}l@{}}
$\Gamma_\text{inhomo., iso.}$\\
$\Gamma_\text{homo., iso.}$\\
$\Gamma_\text{inhomo., exch.}$\\
$\Gamma_\text{homo., exch.}$\\
Exchange Coupling \\
Scaling
\end{tabular}  & \begin{tabular}[t]{@{}l@{}}
0.05(1)~mT\\
0.16(8)~mT\\
1.9(5)~mT\\
1.9(3)~mT\\
24(6)~MHz\\
34(9)\% exch.-coupled
\end{tabular} \\ 
\hline 
\textbf{Single-Crystal} & \begin{tabular}[t]{@{}l@{}}
$\Gamma_\text{inhomo.}$\\
$\Gamma_\text{homo.}$\\
($\phi$,$\theta$,$\psi$) \\
Exchange Coupling \\
Scaling
\end{tabular} & \begin{tabular}[t]{@{}l@{}}
0.08(1)~mT\\
0.115(5)~mT\\
(8.0(1)$^\text{o}$,45.0(6)$^\text{o}$,30.0(1)$^\text{o}$) \\
1.3(0)~MHz \\
37(3)\% 
\end{tabular} \\ 
\hline
\end{tabular} 
\caption{\textbf{Results of fitting each EPR spectrum to a descriptive model with exchange coupling and broadening.} Broadenings are reported as FWHM widths; uncertainties represent 68\% confidence intervals.} 
\label{table:eprfits} 
\end{table}

The weak exchange coupling in the single-crystal sample observed by CW is likely due to a lower nitrogen concentration in the sample. This observation provides supporting evidence for the hypothesis of substantially lower eSD mediated by exchange-coupled clusters of P1 centers. Recent work by Nir-Arad and colleagues has shown these single-crystal HPHT diamonds may have a nitrogen concentration as low as 20~ppm~\cite{nir-aradClusteringHeterogeneousP12025}; much lower than the 110-130~ppm concentration in the powder.

Combining our estimates of the natural and inhomogeneous broadening in each sample, we can then place a bound on the contribution to the broadening due to eSD and therefore the characteristic timescale of eSD itself. Previously, we characterized the two-timescale coherence times at 2.5~GHz for both the powder and single-crystal samples using biexponential decays as 1.9(1)~$\upmu$s and 10(2)~$\upmu$s (powder); and 2.2(5)~$\upmu$s and 18(8)~$\upmu$s (single-crystal)~\cite{shimonLargeRoomTemperature2022}. We note that
\begin{equation}
    \frac{1}{\pi T_\text{SD}}\lesssim\Gamma_\text{homogeneous} - \frac{1}{\pi T_2},
    \label{eq:esd}
\end{equation}
in the case where the expected timescale of eSD, $T_\text{SD}$, is much faster than $T_2$~\cite{abragamElectronParamagneticResonance1970}. For the single-crystal sample this holds; using Eq.~\ref{eq:esd} we find $T_\text{SD}\lesssim0.1$~$\upmu$s. In the powder sample, the exchange-coupled line broadenings are large enough that $T_\text{SD}\lesssim0.006$~$\upmu$s, suggestive of much faster electron-electron spectral diffusion driven by the exchange-coupled clusters in the powder sample. This is consistent with the frequency-modulated DNP results, which suggest intermediate or negligible (slower) eSD in the single-crystal sample but large (fast) eSD in the powder.

\subsection{Truncated cross effect under MW frequency modulation}

We now analyze the central tCE mechanism and hyperfine satellite tCE in the P1 DNP spectrum. The DNP enhancement at the $m_I = 0$ line observed under frequency-modulated MW irradiation, initially attributed to an Overhauser mechanism, is actually driven by a CE-type condition, excepting the pockets of exchange-coupled P1 clusters mixed with isolated P1 centers in the ensemble truncating the polarization efficiency of the match. Though this tCE effect produces an inverted-sign DNP signal at the center of the line, its underlying physics matches that of the tCE. We therefore apply our three-spin (electron-electron-nucleus) model of the tCE under frequency modulation developed in our previous work~\cite{shimonRoomTemperatureDNP2022}. 

The effects of changing $f_m$ followed the results developed for the SE and CE: Just as in the case of the SE and CE, electronic saturation occurs when $f_mT_{1e}>1$, where $T_{1e}$ is the relaxation of the slower-relaxing electron. To activate the tCE condition and properly truncate the efficiency of the CE, $T_{1e,\text{slow}}\geq 100T_{1e,\text{fast}}$. Generally, we find that larger modulation rates can be advantageous for the tCE -- particularly in the case of Figure~\ref{fig:ModMarge}(a). 

\begin{table*}[ht!] 
\centering 
\begin{tabular}{|c|p{4cm}|p{6cm}|p{6cm}} 
\hline 
\textbf{Mechanism} & \textbf{Enhancement Condition} & \textbf{Maximal Enhancement Conditions} \\ 
\hline
\hline 
\textbf{Solid Effect} & Positive or negative enhancement when irradiating at $\omega_e \pm \omega_C$, respectively & Maximal when $\Delta\omega \approx 2\omega_C$ and $f_m T_{1e} \approx 1$ (modulation slow compared to relaxation) \\ 
\hline 
\textbf{Cross Effect} & Occurs when $|\omega_{e1} - \omega_{e2}| = \omega_C$; requires two electrons with distinct resonance frequencies & Maximal when $\Delta\omega \lesssim 2\omega_C$ and electron–electron coupling permits CE matching \\ 
\hline
 \textbf{Truncated Cross Effect} & Effective CE‐type transfer between coupled and isolated P1 centers with $T_{1e,\text{slow}} \gg T_{1e,\text{fast}}$ & Maximal when $\omega_{MW}$ centered near $m_I=0$ peak and $\Delta\omega \gtrsim 2\omega_C$, with $1/T_{1e,\text{slow}} < f_m < 1/T_{1e,\text{fast}}$  \\ 
\hline 
\end{tabular} 
\caption{\textbf{Results of maximal frequency-modulated DNP enhancement for each polarization transfer mechanism at low MW power.}} 
\label{table:modulation_mechanisms} 
\end{table*}

The dynamics of the tCE were studied for four cases of values of $\Delta\omega$: a) no modulation, b) modulation with $\Delta\omega<\omega_C$, c) $\Delta\omega=\omega_C$, and d) $\Delta\omega>\omega_C$~\cite{shimonRoomTemperatureDNP2022}. To examine these cases, we study the effects of eSD. We again compare the behavior in terms of modulation enhancement, the DNP enhancement driven by MW modulation compared to the MCW DNP signal. In the case of negligible eSD, the modulation enhancement grows smaller until the cutoff threshold $\Delta\omega\approx2\omega_C$ is reached, upon which the enhancement begins to sample a large region of the EPR line. Intermediate eSD drives modulation enhancement slowly as $\Delta\omega$ is increased, hitting a plateau at $\Delta\omega\approx2\omega_C$. Large eSD will show a small increase in the tCE modulation enhancement as $\Delta\omega$ increases, but only until the EPR line is fully saturated. 

Our data in the case of the tCE (Figs.~\ref{fig:ModMarge}(a),~\ref{fig:ModMarge}(b), and~\ref{fig:ModBrendan}(a)) generally match these predictions. In Figure~\ref{fig:ModBrendan}(a), for example, tuning the value of $f_m$ between $1/T_{1e,\text{slow}}$ and $1/T_{1e,\text{fast}}$ while pushing $\Delta\omega$ above $2\omega_C$ maximizes modulation enhancement, consistent with a two-timescale electron relaxation model. At small $\Delta\omega$, however, the modulation enhancement drops below one, indicating diminishing DNP signal via frequency modulation until $\Delta\omega$ is greater than $2\omega_C$. This observation is consistent with a model of weaker electron-electron spectral diffusion in the single-crystal sample -- especially compared to the powdered sample measured in our previous work. In the case of CE -- tCE overlap, eSD can further enhance the DNP signal via an indirect CE (iCE)~\cite{hovavEffectsElectronPolarization2015}. We do not observe strong modulation enhancement when the CE and tCE each contribute strong enhancement of the DNP signal. 

\subsection{Enhancing $^{13}$C NMR signals with P1 chirped DNP}

Though the complexities of the DNP spectrum of the P1 center in HPHT diamond, driven by the dynamics of a large ensemble system, make quantitative theoretical analysis challenging, we explore the phenomenological behaviors common to different modulation datasets. We note several phenomena present in many or all of the colormaps shown, and use these to construct a set of conclusions on the efficacy of enhancing DNP signal with frequency-modulated microwaves irradiating P1 centers in diamond. We previously established that typical buildup rates correlate well with the nuclear $T_{1n}$ time in the powder sample, suggesting that we are working in the nonsaturating (linear) microwave regime and that we cannot drive DNP saturation faster than $T_{1n}$~\cite{shimonRoomTemperatureDNP2022}. We observe only a modest difference in $T_{1e}$ in the single-crystal sample as compared to the diamond sample, but note a substantially slower DNP buildup time -- suggesting that though nuclear spin $T_{1n}$ time may be longer in the single-crystal sample than in the powder, the available MW power of 240~mW does not induce a power-saturated DNP spectrum.

We observe that for nearly all of the data shown in Figures~\ref{fig:ModMarge} and~\ref{fig:ModBrendan}, large-amplitude slow modulation tends to minimize enhancement, consistent with powdered diamond under frequency modulation~\cite{shimonRoomTemperatureDNP2022}. In the single-crystal sample, however, the narrower features of the DNP spectrum allow for full inversion of the DNP signal under these modulation conditions. In particular, the SE enhancement tends to drop off or even invert under these conditions. The maximal likelihood of inverted DNP signal when the MW frequency is parked on the SE peak tends to occur for $\Delta\omega\gtrsim2\omega_C$ and $f_m\lesssim1/T_{1e,\text{slow}}$. When multiple SE peaks are available within $\omega_{MW}\pm\Delta\omega = \omega_{MW}\pm\omega_C$, particularly in Figure~\ref{fig:ModMarge}(d), we observe modulation enhancement coinciding with iSE-driven DNP.

Both the SE and CE are susceptible to poor DNP enhancement under large amplitude modulation, particularly at long electron spectral diffusion~\cite{shimonRoomTemperatureDNP2022}. We observe for all DNP mechanisms that the modulation rate $1/f_m$ tends to provide stronger enhancement at small $\Delta\omega$ when it is faster than $T_{1e,\text{fast}}$, because the electron spin does not lose polarization between periods~\cite{hovavDynamicNuclearPolarization2014}. At very fast frequency modulation as compared to $T_{1e}$, however, individual DNP mechanisms never reach saturation fast enough to efficiently transfer electron polarization beyond the equivalent MCW irradiation. Comparing these effects, we note that $1/f_m<<T_{1e}$ yields large modulation enhancement at small $\Delta\omega$ but poor enhancement and occasionally inversion at large $\Delta\omega$, while $1/f_m>>T_{1e}$ yields generally poor modulation enhancement. A summary of the regime in which each mechanism yields maximal modulation enhancement is provided in Table~\ref{table:modulation_mechanisms}.

Finally, we note that frequency-modulated DNP serves as a tool for understanding eSD in single-crystal HPHT diamond. Because the modulation enhancement tends to be maximized using $\Delta\omega\approx2\omega_C$, we suggest that this sample has weak (and perhaps negligible) eSD, in contrast to the powder HPHT diamond sample in our previous work. These observations suggest that frequency modulation can effectively probe eSD strength: strong enhancement near $\Delta\omega\approx2\omega_C$ implies weak eSD, whereas reduced or no enhancement suggests more significant eSD.

\section{Inversion of the DNP signal at long buildup times}
\label{sec:signalinversion}

\begin{figure*}[ht!]
    \centering
   \includegraphics[width=1\linewidth]{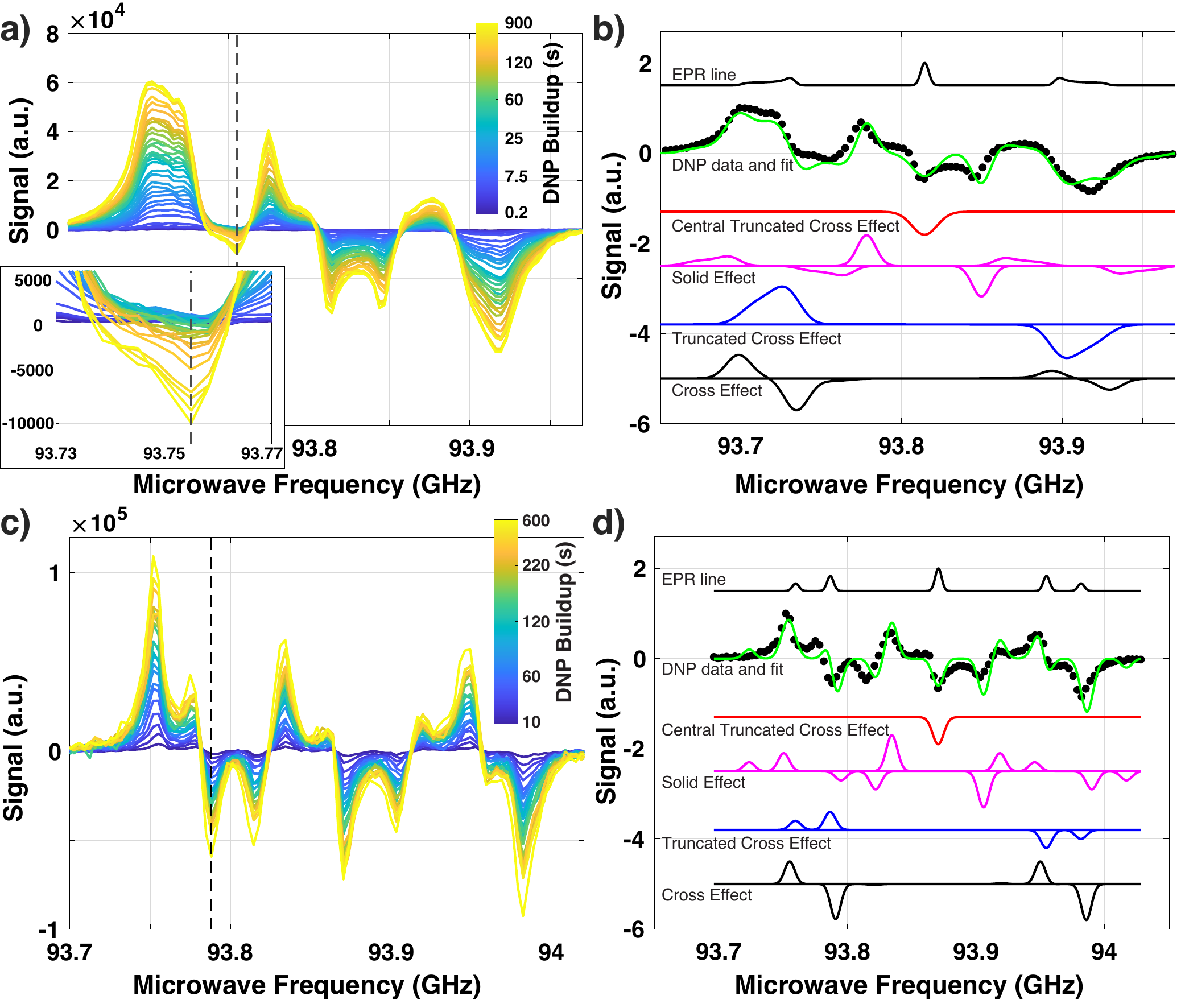}  
\caption{\textbf{MW frequency and buildup time dependence in powdered and single-crystal HPHT diamond at 3.34~T.} (a) DNP spectra for a powder diamond sample with buildup time values ranging from .2 to 900~s. The inset shows the region where the DNP signal inverts sign at long buildup time. The dashed vertical line represents the microwave frequency (93.755~GHz) at which the buildup time dependence is shown in Figure~\ref{fig:3T7Tbuildups}(a). (b) DNP spectral decomposition of the powder sample, at buildup time 900~s. (c) DNP spectra for the single-crystal diamond sample with buildup time values ranging from 10 to 600~s. The dashed vertical line represents the microwave frequency (93.788~GHz) at which the buildup time dependence is shown in Figure~\ref{fig:3T7Tbuildups}(a). (d) DNP spectral decomposition of the single-crystal sample, at buildup time 600~s. In this orientation, similar to Figure~\ref{fig:crystaldecomp}(c), all DNP mechanisms are prominent contributors to the spectrum.}
    \label{fig:3TData_PowSC_Inv}
\end{figure*}

In this section we discuss the use of MCW microwave DNP on the P1 center in a powder sample. The EPR line of the hyperfine satellites in a powder smears the detail of distinct peaks, and for HPHT diamond shows a clear contribution from exchange-coupled P1 clusters, as well~\cite{bussandriP1CenterElectron2024,sternP1CenterNetwork2025}. The DNP spectrum for a powder diamond sample takes a similar form; our previous work shows the same DNP mechanisms (SE, CE, and tCE) are present in the powder~\cite{shimonLargeRoomTemperature2022}. The DNP spectrum of the powder sample is shown in Figure~\ref{fig:3TData_PowSC_Inv}(a). The color scale represents the range of buildup times, from 0.2~s to 900~s. For most microwave frequencies in the spectrum, the DNP-enhanced signal increases monotonically with buildup time. Within a small range of 93.735 - 93.765~GHz, however, the signal enhancement changes sign from positive to negative as buildup time is increased. The inset of (a) focuses on this range of frequencies, showing the effect clearly. 

The DNP spectrum for a single-crystal sample at 3.34~T similarly mirrors the EPR line, as seen in Figs.~\ref{fig:3TData_PowSC_Inv}(c) and~(d). Each DNP mechanism is strongly orientation-dependent due to the Jahn-Teller distortion driving single-axis anisotropy in the P1 electron-nuclear coupling, meaning variations in the orientation of the sample relative to the field allow us to choose which mechanisms are in competition, and the MW frequencies where the competition is observable. 

\begin{figure*}[ht!]
    \centering
\includegraphics[width=1\linewidth]{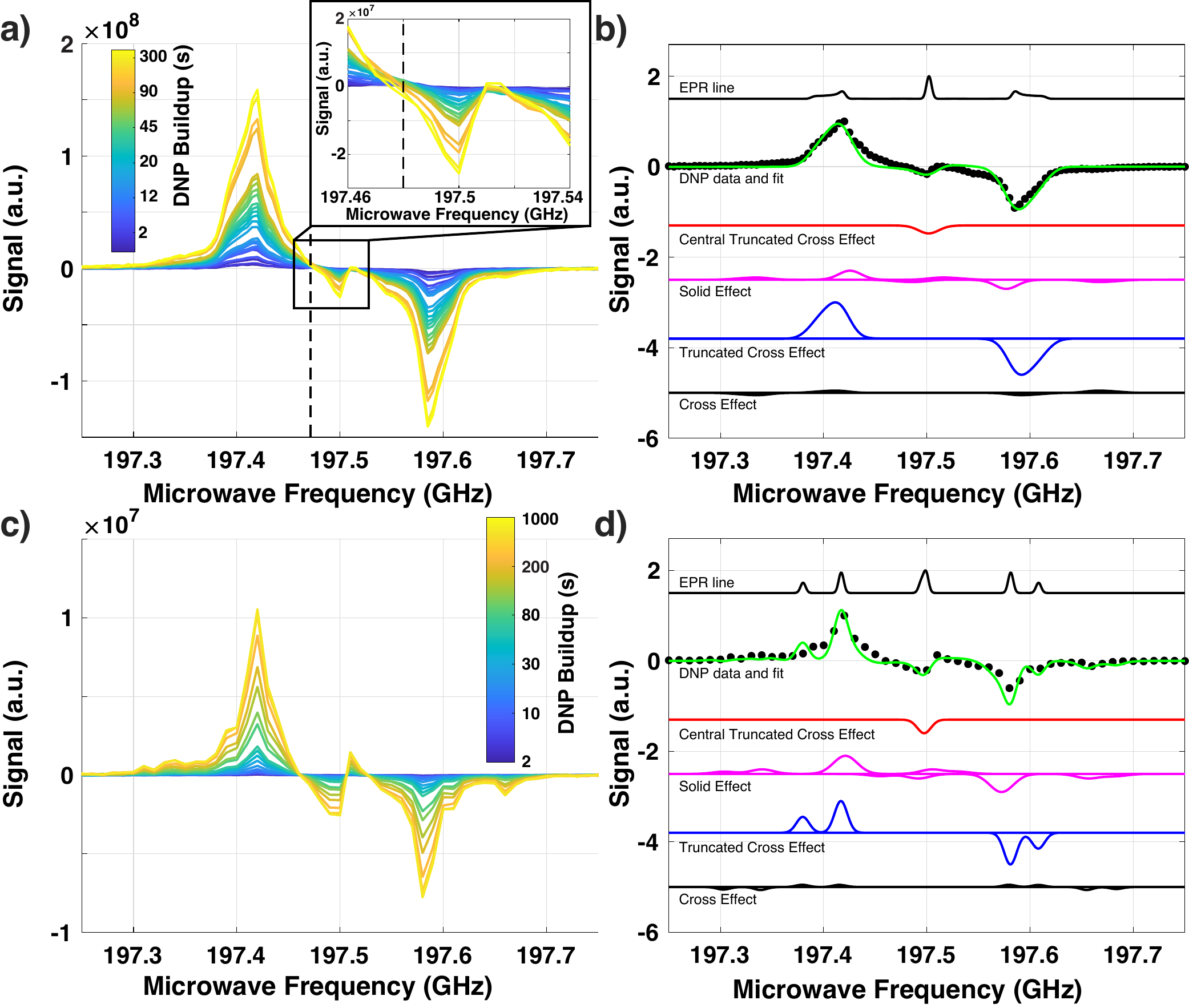}
\caption{\textbf{Powder and single-crystal DNP spectra and corresponding decomposition at 7~T.} (a) DNP spectra for the powder diamond sample with buildup time values ranging from 2 to 300~s. The inset focuses on a range of MW frequencies over which the small signal inversion appears. The dashed vertical line represents the microwave frequency (197.48~GHz) at which the buildup time dependence is shown in Figure~\ref{fig:3T7Tbuildups}(b). b) The decomposition of this DNP spectrum (at 300~s buildup time) shows at least small contributions from all DNP enhancement mechanisms, though the truncated cross effect dominates. (c) DNP spectra for the HPHT single-crystal diamond sample, with buildup time values ranging from 2 to 1000~s. No signal inversion in the single-crystal at 7~T was observed. (d) The decomposition of the single-crystal 7~T DNP spectrum at 1000~s buildup time.}
    \label{fig:7TData_PowSC_Inv}
\end{figure*}

We then examine the same diamond powder and single-crystal samples at 7.05~T with MCW DNP irradiation, as shown in Figure~\ref{fig:7TData_PowSC_Inv}. In panel (a) we observe a small inversion of the signal around 197.5~GHz; panel (b) shows the spectral decomposition of the DNP signal. The diamond powder spectrum is dominated by the truncated cross effect generating positive DNP enhancement at low frequency and negative DNP enhancement at high frequency, in agreement with previous work~\cite{bussandriP1CenterElectron2024}. The single-crystal spectrum at 7~T shows a stronger SE peak from the $m_I = 0$ line than the powder, but remains dominated by the tCE (Fig.~\ref{fig:7TData_PowSC_Inv}(c) and~(d)). The slight $g$-anisotropy resolvable at high field is responsible for the slight anisotropy of the simulated EPR spectra.

\begin{figure}[ht!]
    \centering
\includegraphics[width=1\linewidth]{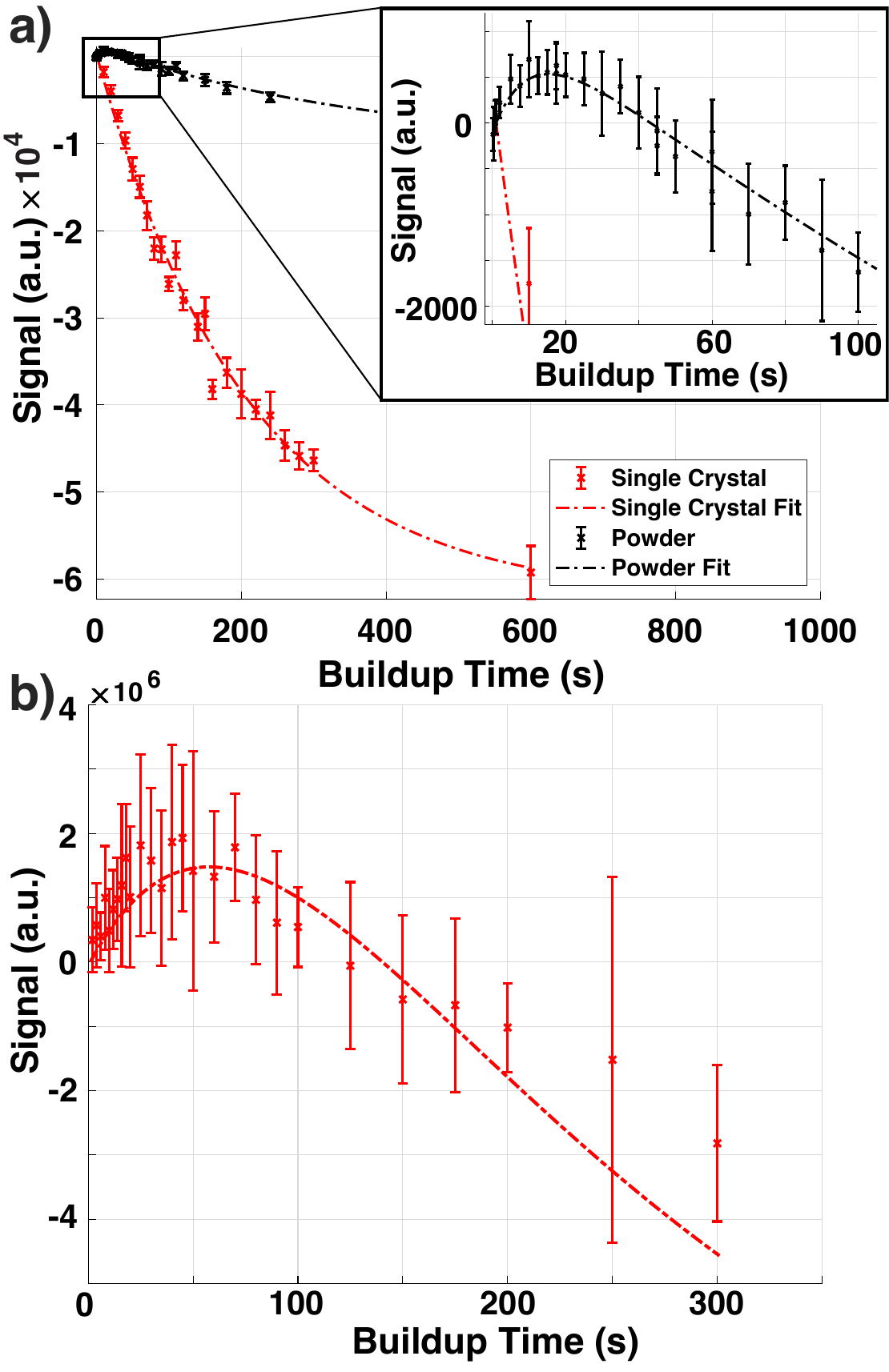}
\caption{\textbf{Buildup curves in the diamond powder emphasize signal inversion.} (a) Buildup curves of the powder (black) and single-crystal (red) diamond samples at 3.34~T show the powder sample has an inversion of the DNP signal as buildup time is increased. The inset shows that positive DNP signal hits a maximum at 15~s. (b) The 7~T powder signal also inverts as buildup time increases. In both panels, the dashed lines show biexponential recovery best-fit lines.}
    \label{fig:3T7Tbuildups}
\end{figure}

Figure~\ref{fig:3T7Tbuildups} shows the buildup time dependence in the powder sample. In panel (a), at 93.755~GHz, the inversion of the DNP signal in the powder sample becomes visible. At short buildup time, the DNP-enhanced NMR signal is positive, and in fact achieves a maximum value at 15~s MW buildup time. At buildup times longer than 15~s, the DNP signal begins to grow smaller; beyond 40~s MW buildup the DNP signal is negative. As will be discussed below, we attribute this effect to a competition between mechanisms of DNP wherein one effect builds up polarization much faster than another. Shown in red is an example of a typical exponential saturation of the DNP signal with no sign inversion -- visible at most frequencies in the powder spectra with DNP signal and at all frequencies in the single-crystal spectra. This example is measured at 93.788~GHz. Figure~\ref{fig:3T7Tbuildups}(b) shows an inversion of DNP signal measured in the powder sample at 7.05~T. In this dataset the inversion is weaker and hits a positive peak at a longer buildup time (roughly 45~s) than the example at 3.34~T; we note that we have lower nominal MW power at 7.05~T than at 3.34~T.

We examined multiple orientations for MW frequencies where competition between DNP mechanisms exists; we have never observed a sign change in the single-crystal buildup curve. Rather, we observed solely monotonic DNP growth (Fig.~\ref{fig:3T7Tbuildups}(a)) at all measured MW frequencies and orientations for the single-crystal sample at both fields. We examined orientations where competition existed between all possible opposite-sign pairs (SE-CE, SE-tCE, and CE-tCE) of mechanisms, and found no signal inversion. This observation is consistent with the hypothesis that the inversion of the buildup curve as MW irradiation time is increased comes from competing mechanisms of DNP driven by different crystallites in the bulk powder. For the case of sign inversion in Figure~\ref{fig:3TData_PowSC_Inv}(a), the competition is likely between the slow-building SE and faster polarization transfer associated with the CE.

\section{Conclusions}
\label{sec:conc}

We present the results of microwave-induced DNP on P1 centers in diamond wherein we demonstrate the suppression of some mechanisms of DNP while enhancing others. We demonstrate this capability in two ways. First, we show that through the use of frequency-modulated microwave irradiation in a single-crystal sample one can use careful choice of modulation parameters to enhance DNP selectively in such a way that the mechanism(s) giving rise to the signal are controlled. Second, we show that even without frequency-modulated microwaves, we observe competing mechanisms of DNP giving rise to a sign change in the DNP spectrum of powdered HPHT diamond.

In the frequency modulation experiment, we optimize the combination of modulation amplitude $\Delta\omega$ and modulation frequency $f_m$ to achieve total DNP modulation enhancement of up to a factor of six, when the central frequency sits at a frequency where the SE, CE, and tCE all compete to give rise a large overall negative MCW DNP enhancement. In another experiment, we demonstrate that varying $\Delta\omega$ and $f_m$ at a central frequency where the SE and CE sum to form a large positive MCW DNP signal gives rise to a modulation enhancement varying from negative three to three. We attribute these effects to the capability of selectively enhancing and suppressing different DNP mechanisms through the use of varying $\Delta\omega$ and $f_m$.

At both fields we observe small signal inversion in a narrow window of the DNP frequency spectrum: short MW buildup times yields a positive DNP signal, while long MW buildup inverts and generates a much larger negative signal. To drive this behavior multiple mechanisms of polarization transfer occurring at the same MW frequency are required. By comparing powder and single-crystal samples, and noting that we observe this effect only in the powder, we conclude that this effect is the result of competition from mechanisms of DNP from different crystallites in the powder; orientational averaging in the powder spectrum thereby produces this DNP turnover effect. 

When examining the inversion of the powder spectrum, we observe a sign change in the MCW DNP spectrum in the frequency region between the $m_I=-1$ and $m_I=0$ hyperfine manifolds at both fields. At steady state the DNP signal is negative; simply by shortening MW buildup time below 40~s we observe positive DNP signal. We examine a decomposition of the spectrum at 15~s and 900~s buildup, noting the contribution of the hyperfine SE peak competing with the tCE and central EPR line positive SE peak is consistent with the sign change in the powder spectrum. The single-electron SE likely builds up faster than the tCE~\cite{nir-aradClusteringHeterogeneousP12025} and creates the short-buildup positive signal, while the tCE dominates at long time. Notably, we do not observe this effect in any single-crystal sample, regardless of the orientation or combination of DNP mechanisms at play. We attribute this sign change to competition between crystallites in the bulk powder. Both systems, the single-crystal diamond and the powder diamond, exhibit regions where different DNP mechanisms can be selectively suppressed or enhanced; we demonstrate this process for both the single-crystal and powder sample.

\section{Materials and methods}
\label{sec:meth}

\subsection{Experiments}

\subsubsection{Sample preparation}

We performed DNP experiments at 3.34~T on both powdered type 1b high-pressure, high-temperature (HPHT) and single-crystal ``stone-cut'' type 1b HPHT diamond samples. The bulk diamond powder was donated by Element 6, and contains microparticles of diameter 15 -- 25 $\upmu$m and are specified to have a nitrogen concentration less than 200~ppm (Element 6 estimated the actual values were about 110 -- 130~ppm). The single-crystal stone-cut HPHT sample was purchased from Element 6 and contains a nitrogen concentration less than 200~ppm. EPR experiments were performed on a Bruker EMX spectrometer.

\subsubsection{DNP spectrometer}

All DNP experiments at 3.34~T were performed on a custom-built W-band DNP spectrometer~\cite{guyDesignCharacterizationWband2015,shimonLargeRoomTemperature2022}. Circuit characterization has shown typical output MW power is 240~mW (24~dBm). The 3.34~T field sets the electron Larmor frequency to 94~GHz and the $^{13}$C Larmor frequency to 35.8~MHz. The 94~GHz electron-irradiating microwave W-band system is described in~\cite{shimonLargeRoomTemperature2022}, and the modulation technique is described in~\cite{shimonRoomTemperatureDNP2022}. NMR experiments at 3.34~T were performed with a SpinCore iSpin (Figs.~\ref{fig:crystaldecomp}(b) and~\ref{fig:ModBrendan}) or a Bruker Avance AQX (Figs.~\ref{fig:crystaldecomp}(c),~\ref{fig:ModMarge},~\ref{fig:3TData_PowSC_Inv}, and~\ref{fig:3T7Tbuildups}); in each case these commercial instruments performed both pulse generation and signal digitization. 7.05~T DNP experiments were performed using a DNP spectrometer with a quasioptical microwave bridge from Bridge12, developed from~\cite{siawVersatileModularQuasi2016}. Nominal power out of the quasioptical bridge is 100~mW. The NMR spectrometer was built around a SpinCore Radioprocessor.

\subsubsection{DNP-NMR experiments}

DNP-NMR was performed with the pulse sequences shown in Figure~\ref{fig:crystaldecomp}(a). Each pulse sequence began with a saturation train of 30 $\uppi/2$-pulses spaced 50~$\upmu$s apart, to crush any remnant magnetization in the sample. The microwaves, which are blanked during both saturation-train pulsing and FID measurement, were then turned on (and chirped, if necessary) via a TTL line from the NMR spectrometer. After the buildup time $\uptau_\text{$\upmu$wave}$ had elapsed, single-pulse FID readout via a 10 $\upmu$s long $\uppi/2$ pulse was performed. The $\uppi/2$ pulse length was calibrated with a single-pulse Rabi nutation experiment, such that the pulse length generating maximum FID signal was chosen.

\subsubsection{Data processing}

Data processing was performed in MATLAB using custom software. In all cases, two points of left-shift were used. The data were then baseline-corrected, phase-corrected, and filtered with 200~Hz exponential line broadening. We show the integrated intensity of the Fourier-transformed FID at $^{13}$C resonance. To calculate modulation enhancement, we divide the frequency-modulated DNP signal by the MCW DNP signal. No enhancement is therefore equal to one; inverted signal is negative. 

\subsection{Simulations}

\subsubsection{Simulating the EPR line}

EPR simulations and fitting was performed using least-squares regression of spectra simulated using EasySpin~\cite{stollEasySpinComprehensiveSoftware2006}. For the DNP simulations, the EPR line of the P1 center (described by Eq.~\ref{eq:spinham}) forms the basis for DNP spectrum simulations. Fitting parameters for each DNP spectrum were the relative amplitudes of each mechanism, convolution linewidths for the SE and CE/tCE, and (for the single-crystal spectra) Euler angles. The single-crystal sample is a twinned crystal, but this only visible at certain orientations -- in our datatsets, twinning only appears in the EPR data. Simulations were performed with EasySpin, as described in our previous publication~\cite{shimonLargeRoomTemperature2022}.

\subsubsection{Fitting DNP spectra}

DNP spectrum decompositions into component mechanisms were performed through a least-squares fit in Matlab, wherein the relative magnitude of each mechanism for each EPR-line manifold was a fitting parameter, as well as Gaussian convolution FWHM for each line. 

\section*{Author Declarations}
\subsection*{Conflict of Interest}
The authors declare that they have no known competing financial interests or personal relationships that could influence or appear to influence work reported in this paper.
\newline
\subsection*{Data Availability}
The data that support the findings of this study are available from the corresponding author upon reasonable request.
\newline
\subsection*{Acknowledgments}
We thank Element 6 for the donation of the powder sample used in these experiments. We thank J.~Logan, S.~Ganguly, G.~K.~Kurian, and L.~Joseph for useful conversations and advice, P.~Defino for assistance with and usage of the Bruker EMX for X-band EPR, and J.~Bonneau and T.~Maly for assistance with assembling the microwave bridge and NMR spectrometer used at 7~T. This work was supported by the National Science Foundation under grant \#2203681.

\bibliography{OverlapMech}

\begin{thebibliography}{39}%
\makeatletter
\providecommand \@ifxundefined [1]{%
 \@ifx{#1\undefined}
}%
\providecommand \@ifnum [1]{%
 \ifnum #1\expandafter \@firstoftwo
 \else \expandafter \@secondoftwo
 \fi
}%
\providecommand \@ifx [1]{%
 \ifx #1\expandafter \@firstoftwo
 \else \expandafter \@secondoftwo
 \fi
}%
\providecommand \natexlab [1]{#1}%
\providecommand \enquote  [1]{``#1''}%
\providecommand \bibnamefont  [1]{#1}%
\providecommand \bibfnamefont [1]{#1}%
\providecommand \citenamefont [1]{#1}%
\providecommand \href@noop [0]{\@secondoftwo}%
\providecommand \href [0]{\begingroup \@sanitize@url \@href}%
\providecommand \@href[1]{\@@startlink{#1}\@@href}%
\providecommand \@@href[1]{\endgroup#1\@@endlink}%
\providecommand \@sanitize@url [0]{\catcode `\\12\catcode `\$12\catcode `\&12\catcode `\#12\catcode `\^12\catcode `\_12\catcode `\%12\relax}%
\providecommand \@@startlink[1]{}%
\providecommand \@@endlink[0]{}%
\providecommand \url  [0]{\begingroup\@sanitize@url \@url }%
\providecommand \@url [1]{\endgroup\@href {#1}{\urlprefix }}%
\providecommand \urlprefix  [0]{URL }%
\providecommand \Eprint [0]{\href }%
\providecommand \doibase [0]{https://doi.org/}%
\providecommand \selectlanguage [0]{\@gobble}%
\providecommand \bibinfo  [0]{\@secondoftwo}%
\providecommand \bibfield  [0]{\@secondoftwo}%
\providecommand \translation [1]{[#1]}%
\providecommand \BibitemOpen [0]{}%
\providecommand \bibitemStop [0]{}%
\providecommand \bibitemNoStop [0]{.\EOS\space}%
\providecommand \EOS [0]{\spacefactor3000\relax}%
\providecommand \BibitemShut  [1]{\csname bibitem#1\endcsname}%
\let\auto@bib@innerbib\@empty
\bibitem [{\citenamefont {King}\ \emph {et~al.}(2010)\citenamefont {King}, \citenamefont {Coles},\ and\ \citenamefont {Reimer}}]{kingOpticalPolarization13C2010}%
  \BibitemOpen
  \bibfield  {author} {\bibinfo {author} {\bibfnamefont {J.~P.}\ \bibnamefont {King}}, \bibinfo {author} {\bibfnamefont {P.~J.}\ \bibnamefont {Coles}},\ and\ \bibinfo {author} {\bibfnamefont {J.~A.}\ \bibnamefont {Reimer}},\ }\bibfield  {title} {\bibinfo {title} {Optical polarization of {\textsuperscript{13}}{{C}} nuclei in diamond through nitrogen vacancy centers},\ }\href {https://doi.org/10.1103/PhysRevB.81.073201} {\bibfield  {journal} {\bibinfo  {journal} {Phys. Rev. B}\ }\textbf {\bibinfo {volume} {81}},\ \bibinfo {pages} {073201} (\bibinfo {year} {2010})}\BibitemShut {NoStop}%
\bibitem [{\citenamefont {King}\ \emph {et~al.}(2015)\citenamefont {King}, \citenamefont {Jeong}, \citenamefont {Vassiliou}, \citenamefont {Shin}, \citenamefont {Page}, \citenamefont {Avalos}, \citenamefont {Wang},\ and\ \citenamefont {Pines}}]{kingRoomtemperatureSituNuclear2015}%
  \BibitemOpen
  \bibfield  {author} {\bibinfo {author} {\bibfnamefont {J.~P.}\ \bibnamefont {King}}, \bibinfo {author} {\bibfnamefont {K.}~\bibnamefont {Jeong}}, \bibinfo {author} {\bibfnamefont {C.~C.}\ \bibnamefont {Vassiliou}}, \bibinfo {author} {\bibfnamefont {C.~S.}\ \bibnamefont {Shin}}, \bibinfo {author} {\bibfnamefont {R.~H.}\ \bibnamefont {Page}}, \bibinfo {author} {\bibfnamefont {C.~E.}\ \bibnamefont {Avalos}}, \bibinfo {author} {\bibfnamefont {H.-J.}\ \bibnamefont {Wang}},\ and\ \bibinfo {author} {\bibfnamefont {A.}~\bibnamefont {Pines}},\ }\bibfield  {title} {\bibinfo {title} {Room-temperature in situ nuclear spin hyperpolarization from optically pumped nitrogen vacancy centres in diamond},\ }\href {https://doi.org/10.1038/ncomms9965} {\bibfield  {journal} {\bibinfo  {journal} {Nat Commun}\ }\textbf {\bibinfo {volume} {6}},\ \bibinfo {pages} {8965} (\bibinfo {year} {2015})}\BibitemShut {NoStop}%
\bibitem [{\citenamefont {Ajoy}\ \emph {et~al.}(2018)\citenamefont {Ajoy}, \citenamefont {Nazaryan}, \citenamefont {Liu}, \citenamefont {Lv}, \citenamefont {Safvati}, \citenamefont {Wang}, \citenamefont {Druga}, \citenamefont {Reimer}, \citenamefont {Suter}, \citenamefont {Ramanathan}, \citenamefont {Meriles},\ and\ \citenamefont {Pines}}]{ajoyEnhancedDynamicNuclear2018}%
  \BibitemOpen
  \bibfield  {author} {\bibinfo {author} {\bibfnamefont {A.}~\bibnamefont {Ajoy}}, \bibinfo {author} {\bibfnamefont {R.}~\bibnamefont {Nazaryan}}, \bibinfo {author} {\bibfnamefont {K.}~\bibnamefont {Liu}}, \bibinfo {author} {\bibfnamefont {X.}~\bibnamefont {Lv}}, \bibinfo {author} {\bibfnamefont {B.}~\bibnamefont {Safvati}}, \bibinfo {author} {\bibfnamefont {G.}~\bibnamefont {Wang}}, \bibinfo {author} {\bibfnamefont {E.}~\bibnamefont {Druga}}, \bibinfo {author} {\bibfnamefont {J.~A.}\ \bibnamefont {Reimer}}, \bibinfo {author} {\bibfnamefont {D.}~\bibnamefont {Suter}}, \bibinfo {author} {\bibfnamefont {C.}~\bibnamefont {Ramanathan}}, \bibinfo {author} {\bibfnamefont {C.~A.}\ \bibnamefont {Meriles}},\ and\ \bibinfo {author} {\bibfnamefont {A.}~\bibnamefont {Pines}},\ }\bibfield  {title} {\bibinfo {title} {Enhanced dynamic nuclear polarization via swept microwave frequency combs},\ }\href {https://doi.org/10.1073/pnas.1807125115} {\bibfield  {journal} {\bibinfo  {journal} {Proc. Natl. Acad. Sci. U.S.A.}\
  }\textbf {\bibinfo {volume} {115}},\ \bibinfo {pages} {10576} (\bibinfo {year} {2018})}\BibitemShut {NoStop}%
\bibitem [{\citenamefont {Ajoy}\ \emph {et~al.}(2019)\citenamefont {Ajoy}, \citenamefont {Safvati}, \citenamefont {Nazaryan}, \citenamefont {Oon}, \citenamefont {Han}, \citenamefont {Raghavan}, \citenamefont {Nirodi}, \citenamefont {Aguilar}, \citenamefont {Liu}, \citenamefont {Cai}, \citenamefont {Lv}, \citenamefont {Druga}, \citenamefont {Ramanathan}, \citenamefont {Reimer}, \citenamefont {Meriles}, \citenamefont {Suter},\ and\ \citenamefont {Pines}}]{ajoyHyperpolarizedRelaxometryBased2019}%
  \BibitemOpen
  \bibfield  {author} {\bibinfo {author} {\bibfnamefont {A.}~\bibnamefont {Ajoy}}, \bibinfo {author} {\bibfnamefont {B.}~\bibnamefont {Safvati}}, \bibinfo {author} {\bibfnamefont {R.}~\bibnamefont {Nazaryan}}, \bibinfo {author} {\bibfnamefont {J.~T.}\ \bibnamefont {Oon}}, \bibinfo {author} {\bibfnamefont {B.}~\bibnamefont {Han}}, \bibinfo {author} {\bibfnamefont {P.}~\bibnamefont {Raghavan}}, \bibinfo {author} {\bibfnamefont {R.}~\bibnamefont {Nirodi}}, \bibinfo {author} {\bibfnamefont {A.}~\bibnamefont {Aguilar}}, \bibinfo {author} {\bibfnamefont {K.}~\bibnamefont {Liu}}, \bibinfo {author} {\bibfnamefont {X.}~\bibnamefont {Cai}}, \bibinfo {author} {\bibfnamefont {X.}~\bibnamefont {Lv}}, \bibinfo {author} {\bibfnamefont {E.}~\bibnamefont {Druga}}, \bibinfo {author} {\bibfnamefont {C.}~\bibnamefont {Ramanathan}}, \bibinfo {author} {\bibfnamefont {J.~A.}\ \bibnamefont {Reimer}}, \bibinfo {author} {\bibfnamefont {C.~A.}\ \bibnamefont {Meriles}}, \bibinfo {author} {\bibfnamefont {D.}~\bibnamefont {Suter}},\ and\
  \bibinfo {author} {\bibfnamefont {A.}~\bibnamefont {Pines}},\ }\bibfield  {title} {\bibinfo {title} {Hyperpolarized relaxometry based nuclear {{T}}{\textsubscript{1}} noise spectroscopy in diamond},\ }\href {https://doi.org/10.1038/s41467-019-13042-3} {\bibfield  {journal} {\bibinfo  {journal} {Nat Commun}\ }\textbf {\bibinfo {volume} {10}},\ \bibinfo {pages} {5160} (\bibinfo {year} {2019})}\BibitemShut {NoStop}%
\bibitem [{\citenamefont {Ajoy}\ \emph {et~al.}(2021)\citenamefont {Ajoy}, \citenamefont {Sarkar}, \citenamefont {Druga}, \citenamefont {Zangara}, \citenamefont {Pagliero}, \citenamefont {Meriles},\ and\ \citenamefont {Reimer}}]{ajoyLowfieldMicrowavemediatedOptical2021}%
  \BibitemOpen
  \bibfield  {author} {\bibinfo {author} {\bibfnamefont {A.}~\bibnamefont {Ajoy}}, \bibinfo {author} {\bibfnamefont {A.}~\bibnamefont {Sarkar}}, \bibinfo {author} {\bibfnamefont {E.}~\bibnamefont {Druga}}, \bibinfo {author} {\bibfnamefont {P.}~\bibnamefont {Zangara}}, \bibinfo {author} {\bibfnamefont {D.}~\bibnamefont {Pagliero}}, \bibinfo {author} {\bibfnamefont {C.}~\bibnamefont {Meriles}},\ and\ \bibinfo {author} {\bibfnamefont {J.}~\bibnamefont {Reimer}},\ }\bibfield  {title} {\bibinfo {title} {Low-field microwave-mediated optical hyperpolarization in optically pumped diamond},\ }\href {https://doi.org/10.1016/j.jmr.2021.107021} {\bibfield  {journal} {\bibinfo  {journal} {Journal of Magnetic Resonance}\ }\textbf {\bibinfo {volume} {331}},\ \bibinfo {pages} {107021} (\bibinfo {year} {2021})}\BibitemShut {NoStop}%
\bibitem [{\citenamefont {Scott}\ \emph {et~al.}(2016)\citenamefont {Scott}, \citenamefont {Drake},\ and\ \citenamefont {Reimer}}]{scottPhenomenologyOpticallyPumped2016}%
  \BibitemOpen
  \bibfield  {author} {\bibinfo {author} {\bibfnamefont {E.}~\bibnamefont {Scott}}, \bibinfo {author} {\bibfnamefont {M.}~\bibnamefont {Drake}},\ and\ \bibinfo {author} {\bibfnamefont {J.~A.}\ \bibnamefont {Reimer}},\ }\bibfield  {title} {\bibinfo {title} {The phenomenology of optically pumped {\textsuperscript{13}}{{C NMR}} in diamond at 7.05 {{T}}: {{Room}} temperature polarization, orientation dependence, and the effect of defect concentration on polarization dynamics},\ }\href {https://doi.org/10.1016/j.jmr.2016.01.001} {\bibfield  {journal} {\bibinfo  {journal} {Journal of Magnetic Resonance}\ }\textbf {\bibinfo {volume} {264}},\ \bibinfo {pages} {154} (\bibinfo {year} {2016})}\BibitemShut {NoStop}%
\bibitem [{\citenamefont {Reynhardt}\ and\ \citenamefont {High}(1998)}]{reynhardtDynamicNuclearPolarization1998}%
  \BibitemOpen
  \bibfield  {author} {\bibinfo {author} {\bibfnamefont {E.~C.}\ \bibnamefont {Reynhardt}}\ and\ \bibinfo {author} {\bibfnamefont {G.~L.}\ \bibnamefont {High}},\ }\bibfield  {title} {\bibinfo {title} {Dynamic nuclear polarization of diamond. {{I}}. {{Solid}} state and thermal mixing effects},\ }\href {https://doi.org/10.1063/1.477009} {\bibfield  {journal} {\bibinfo  {journal} {The Journal of Chemical Physics}\ }\textbf {\bibinfo {volume} {109}},\ \bibinfo {pages} {4090} (\bibinfo {year} {1998})}\BibitemShut {NoStop}%
\bibitem [{\citenamefont {Cox}\ \emph {et~al.}(1994)\citenamefont {Cox}, \citenamefont {Newton},\ and\ \citenamefont {Baker}}]{cox13C14N15N1994}%
  \BibitemOpen
  \bibfield  {author} {\bibinfo {author} {\bibfnamefont {A.}~\bibnamefont {Cox}}, \bibinfo {author} {\bibfnamefont {M.~E.}\ \bibnamefont {Newton}},\ and\ \bibinfo {author} {\bibfnamefont {J.~M.}\ \bibnamefont {Baker}},\ }\bibfield  {title} {\bibinfo {title} {{\textsuperscript{13}}{{C}}, {\textsuperscript{14}}{{N}} and {\textsuperscript{15}}{{N ENDOR}} measurements on the single substitutional nitrogen centre ({{P1}}) in diamond},\ }\href {https://doi.org/10.1088/0953-8984/6/2/025} {\bibfield  {journal} {\bibinfo  {journal} {J. Phys.: Condens. Matter}\ }\textbf {\bibinfo {volume} {6}},\ \bibinfo {pages} {551} (\bibinfo {year} {1994})}\BibitemShut {NoStop}%
\bibitem [{\citenamefont {Takahashi}\ \emph {et~al.}(2008)\citenamefont {Takahashi}, \citenamefont {Hanson}, \citenamefont {Van~Tol}, \citenamefont {Sherwin},\ and\ \citenamefont {Awschalom}}]{takahashiQuenchingSpinDecoherence2008}%
  \BibitemOpen
  \bibfield  {author} {\bibinfo {author} {\bibfnamefont {S.}~\bibnamefont {Takahashi}}, \bibinfo {author} {\bibfnamefont {R.}~\bibnamefont {Hanson}}, \bibinfo {author} {\bibfnamefont {J.}~\bibnamefont {Van~Tol}}, \bibinfo {author} {\bibfnamefont {M.~S.}\ \bibnamefont {Sherwin}},\ and\ \bibinfo {author} {\bibfnamefont {D.~D.}\ \bibnamefont {Awschalom}},\ }\bibfield  {title} {\bibinfo {title} {Quenching {{Spin Decoherence}} in {{Diamond}} through {{Spin Bath Polarization}}},\ }\href {https://doi.org/10.1103/PhysRevLett.101.047601} {\bibfield  {journal} {\bibinfo  {journal} {Phys. Rev. Lett.}\ }\textbf {\bibinfo {volume} {101}},\ \bibinfo {pages} {047601} (\bibinfo {year} {2008})}\BibitemShut {NoStop}%
\bibitem [{\citenamefont {Casabianca}\ \emph {et~al.}(2011)\citenamefont {Casabianca}, \citenamefont {Shames}, \citenamefont {Panich}, \citenamefont {Shenderova},\ and\ \citenamefont {Frydman}}]{casabiancaFactorsAffectingDNP2011}%
  \BibitemOpen
  \bibfield  {author} {\bibinfo {author} {\bibfnamefont {L.~B.}\ \bibnamefont {Casabianca}}, \bibinfo {author} {\bibfnamefont {A.~I.}\ \bibnamefont {Shames}}, \bibinfo {author} {\bibfnamefont {A.~M.}\ \bibnamefont {Panich}}, \bibinfo {author} {\bibfnamefont {O.}~\bibnamefont {Shenderova}},\ and\ \bibinfo {author} {\bibfnamefont {L.}~\bibnamefont {Frydman}},\ }\bibfield  {title} {\bibinfo {title} {Factors {{Affecting DNP NMR}} in {{Polycrystalline Diamond Samples}}},\ }\href {https://doi.org/10.1021/jp206167j} {\bibfield  {journal} {\bibinfo  {journal} {J. Phys. Chem. C}\ }\textbf {\bibinfo {volume} {115}},\ \bibinfo {pages} {19041} (\bibinfo {year} {2011})}\BibitemShut {NoStop}%
\bibitem [{\citenamefont {Von~Witte}\ \emph {et~al.}(2025)\citenamefont {Von~Witte}, \citenamefont {Himmler}, \citenamefont {Tamarov}, \citenamefont {Moilanen}, \citenamefont {Ernst},\ and\ \citenamefont {Kozerke}}]{vonwitteTemperatureDependentDynamicNuclear2025}%
  \BibitemOpen
  \bibfield  {author} {\bibinfo {author} {\bibfnamefont {G.}~\bibnamefont {Von~Witte}}, \bibinfo {author} {\bibfnamefont {A.}~\bibnamefont {Himmler}}, \bibinfo {author} {\bibfnamefont {K.}~\bibnamefont {Tamarov}}, \bibinfo {author} {\bibfnamefont {J.~O.}\ \bibnamefont {Moilanen}}, \bibinfo {author} {\bibfnamefont {M.}~\bibnamefont {Ernst}},\ and\ \bibinfo {author} {\bibfnamefont {S.}~\bibnamefont {Kozerke}},\ }\bibfield  {title} {\bibinfo {title} {Temperature-{{Dependent Dynamic Nuclear Polarization}} of {{Diamond}}},\ }\href {https://doi.org/10.1021/acs.jpcc.5c02747} {\bibfield  {journal} {\bibinfo  {journal} {J. Phys. Chem. C}\ }\textbf {\bibinfo {volume} {129}},\ \bibinfo {pages} {12577} (\bibinfo {year} {2025})}\BibitemShut {NoStop}%
\bibitem [{\citenamefont {Shimon}\ \emph {et~al.}(2022{\natexlab{a}})\citenamefont {Shimon}, \citenamefont {Cantwell}, \citenamefont {Joseph}, \citenamefont {Williams}, \citenamefont {Peng}, \citenamefont {Takahashi},\ and\ \citenamefont {Ramanathan}}]{shimonLargeRoomTemperature2022}%
  \BibitemOpen
  \bibfield  {author} {\bibinfo {author} {\bibfnamefont {D.}~\bibnamefont {Shimon}}, \bibinfo {author} {\bibfnamefont {K.~A.}\ \bibnamefont {Cantwell}}, \bibinfo {author} {\bibfnamefont {L.}~\bibnamefont {Joseph}}, \bibinfo {author} {\bibfnamefont {E.~Q.}\ \bibnamefont {Williams}}, \bibinfo {author} {\bibfnamefont {Z.}~\bibnamefont {Peng}}, \bibinfo {author} {\bibfnamefont {S.}~\bibnamefont {Takahashi}},\ and\ \bibinfo {author} {\bibfnamefont {C.}~\bibnamefont {Ramanathan}},\ }\bibfield  {title} {\bibinfo {title} {Large room temperature bulk {{DNP}} of {\textsuperscript{13}}{{C}} via {{P1}} centers in diamond},\ }\href {https://doi.org/10.1021/acs.jpcc.2c06145} {\bibfield  {journal} {\bibinfo  {journal} {J. Phys. Chem. C}\ }\textbf {\bibinfo {volume} {126}},\ \bibinfo {pages} {17777} (\bibinfo {year} {2022}{\natexlab{a}})}\BibitemShut {NoStop}%
\bibitem [{\citenamefont {Shimon}\ \emph {et~al.}(2022{\natexlab{b}})\citenamefont {Shimon}, \citenamefont {Cantwell}, \citenamefont {Joseph},\ and\ \citenamefont {Ramanathan}}]{shimonRoomTemperatureDNP2022}%
  \BibitemOpen
  \bibfield  {author} {\bibinfo {author} {\bibfnamefont {D.}~\bibnamefont {Shimon}}, \bibinfo {author} {\bibfnamefont {K.}~\bibnamefont {Cantwell}}, \bibinfo {author} {\bibfnamefont {L.}~\bibnamefont {Joseph}},\ and\ \bibinfo {author} {\bibfnamefont {C.}~\bibnamefont {Ramanathan}},\ }\bibfield  {title} {\bibinfo {title} {Room temperature {{DNP}} of diamond powder using frequency modulation},\ }\href {https://doi.org/10.1016/j.ssnmr.2022.101833} {\bibfield  {journal} {\bibinfo  {journal} {Solid State Nuclear Magnetic Resonance}\ }\textbf {\bibinfo {volume} {122}},\ \bibinfo {pages} {101833} (\bibinfo {year} {2022}{\natexlab{b}})}\BibitemShut {NoStop}%
\bibitem [{\citenamefont {Bussandri}\ \emph {et~al.}(2024)\citenamefont {Bussandri}, \citenamefont {Shimon}, \citenamefont {Equbal}, \citenamefont {Ren}, \citenamefont {Takahashi}, \citenamefont {Ramanathan},\ and\ \citenamefont {Han}}]{bussandriP1CenterElectron2024}%
  \BibitemOpen
  \bibfield  {author} {\bibinfo {author} {\bibfnamefont {S.}~\bibnamefont {Bussandri}}, \bibinfo {author} {\bibfnamefont {D.}~\bibnamefont {Shimon}}, \bibinfo {author} {\bibfnamefont {A.}~\bibnamefont {Equbal}}, \bibinfo {author} {\bibfnamefont {Y.}~\bibnamefont {Ren}}, \bibinfo {author} {\bibfnamefont {S.}~\bibnamefont {Takahashi}}, \bibinfo {author} {\bibfnamefont {C.}~\bibnamefont {Ramanathan}},\ and\ \bibinfo {author} {\bibfnamefont {S.}~\bibnamefont {Han}},\ }\bibfield  {title} {\bibinfo {title} {P1 center electron spin clusters are prevalent in type {{Ib}} diamonds},\ }\href {https://doi.org/10.1021/jacs.3c06705} {\bibfield  {journal} {\bibinfo  {journal} {J. Am. Chem. Soc.}\ }\textbf {\bibinfo {volume} {146}},\ \bibinfo {pages} {5088} (\bibinfo {year} {2024})}\BibitemShut {NoStop}%
\bibitem [{\citenamefont {{Nir-Arad}}\ \emph {et~al.}(2024{\natexlab{a}})\citenamefont {{Nir-Arad}}, \citenamefont {Laster}, \citenamefont {Daksi}, \citenamefont {Manukovsky},\ and\ \citenamefont {Kaminker}}]{nir-aradPeculiarEPRSpectra2024}%
  \BibitemOpen
  \bibfield  {author} {\bibinfo {author} {\bibfnamefont {O.}~\bibnamefont {{Nir-Arad}}}, \bibinfo {author} {\bibfnamefont {E.}~\bibnamefont {Laster}}, \bibinfo {author} {\bibfnamefont {M.}~\bibnamefont {Daksi}}, \bibinfo {author} {\bibfnamefont {N.}~\bibnamefont {Manukovsky}},\ and\ \bibinfo {author} {\bibfnamefont {I.}~\bibnamefont {Kaminker}},\ }\bibfield  {title} {\bibinfo {title} {On the peculiar {{EPR}} spectra of {{P1}} centers at high (12--20 {{T}}) magnetic fields},\ }\href {https://doi.org/10.1039/d4cp03055a} {\bibfield  {journal} {\bibinfo  {journal} {Phys. Chem. Chem. Phys.}\ }\textbf {\bibinfo {volume} {26}},\ \bibinfo {pages} {27633} (\bibinfo {year} {2024}{\natexlab{a}})}\BibitemShut {NoStop}%
\bibitem [{\citenamefont {{Nir-Arad}}\ \emph {et~al.}(2024{\natexlab{b}})\citenamefont {{Nir-Arad}}, \citenamefont {Shlomi}, \citenamefont {Manukovsky}, \citenamefont {Laster},\ and\ \citenamefont {Kaminker}}]{nir-aradNitrogenSubstitutionsAggregation2024}%
  \BibitemOpen
  \bibfield  {author} {\bibinfo {author} {\bibfnamefont {O.}~\bibnamefont {{Nir-Arad}}}, \bibinfo {author} {\bibfnamefont {D.~H.}\ \bibnamefont {Shlomi}}, \bibinfo {author} {\bibfnamefont {N.}~\bibnamefont {Manukovsky}}, \bibinfo {author} {\bibfnamefont {E.}~\bibnamefont {Laster}},\ and\ \bibinfo {author} {\bibfnamefont {I.}~\bibnamefont {Kaminker}},\ }\bibfield  {title} {\bibinfo {title} {Nitrogen {{Substitutions Aggregation}} and {{Clustering}} in {{Diamonds}} as {{Revealed}} by {{High-Field Electron Paramagnetic Resonance}}},\ }\href {https://doi.org/10.1021/jacs.3c06739} {\bibfield  {journal} {\bibinfo  {journal} {J. Am. Chem. Soc.}\ }\textbf {\bibinfo {volume} {146}},\ \bibinfo {pages} {5100} (\bibinfo {year} {2024}{\natexlab{b}})}\BibitemShut {NoStop}%
\bibitem [{\citenamefont {{Nir-Arad}}\ \emph {et~al.}(2025)\citenamefont {{Nir-Arad}}, \citenamefont {Shlomi}, \citenamefont {Chaklashiya}, \citenamefont {Manukovsky},\ and\ \citenamefont {Kaminker}}]{nir-aradClusteringHeterogeneousP12025}%
  \BibitemOpen
  \bibfield  {author} {\bibinfo {author} {\bibfnamefont {O.}~\bibnamefont {{Nir-Arad}}}, \bibinfo {author} {\bibfnamefont {D.~H.}\ \bibnamefont {Shlomi}}, \bibinfo {author} {\bibfnamefont {R.~K.}\ \bibnamefont {Chaklashiya}}, \bibinfo {author} {\bibfnamefont {N.}~\bibnamefont {Manukovsky}},\ and\ \bibinfo {author} {\bibfnamefont {I.}~\bibnamefont {Kaminker}},\ }\bibfield  {title} {\bibinfo {title} {Clustering and {{Heterogeneous P1 Distributions}} in {{Diamond Govern DNP Mechanisms}} at 6.9 and 13.8 {{T}}},\ }\href {https://doi.org/10.1021/acs.jpclett.5c01809} {\bibfield  {journal} {\bibinfo  {journal} {J. Phys. Chem. Lett.}\ }\textbf {\bibinfo {volume} {16}},\ \bibinfo {pages} {10952} (\bibinfo {year} {2025})}\BibitemShut {NoStop}%
\bibitem [{\citenamefont {Stern}\ \emph {et~al.}(2025)\citenamefont {Stern}, \citenamefont {Cui}, \citenamefont {Chaklashiya}, \citenamefont {Tobar}, \citenamefont {Judd}, \citenamefont {{Nir-Arad}}, \citenamefont {Shimon}, \citenamefont {Kaminker}, \citenamefont {Takahashi}, \citenamefont {Sirigiri},\ and\ \citenamefont {Han}}]{sternP1CenterNetwork2025}%
  \BibitemOpen
  \bibfield  {author} {\bibinfo {author} {\bibfnamefont {Q.}~\bibnamefont {Stern}}, \bibinfo {author} {\bibfnamefont {J.}~\bibnamefont {Cui}}, \bibinfo {author} {\bibfnamefont {R.}~\bibnamefont {Chaklashiya}}, \bibinfo {author} {\bibfnamefont {C.}~\bibnamefont {Tobar}}, \bibinfo {author} {\bibfnamefont {M.}~\bibnamefont {Judd}}, \bibinfo {author} {\bibfnamefont {O.}~\bibnamefont {{Nir-Arad}}}, \bibinfo {author} {\bibfnamefont {D.}~\bibnamefont {Shimon}}, \bibinfo {author} {\bibfnamefont {I.}~\bibnamefont {Kaminker}}, \bibinfo {author} {\bibfnamefont {H.}~\bibnamefont {Takahashi}}, \bibinfo {author} {\bibfnamefont {J.~R.}\ \bibnamefont {Sirigiri}},\ and\ \bibinfo {author} {\bibfnamefont {S.}~\bibnamefont {Han}},\ }\href {https://doi.org/10.26434/chemrxiv-2025-3r9qt} {\bibinfo {title} {P1 center network in high-pressure high-temperature diamonds is a readily accessible source of nuclear hyperpolarization at 14 {{T}}}} (\bibinfo {year} {2025})\BibitemShut {NoStop}%
\bibitem [{\citenamefont {Stoll}\ and\ \citenamefont {Schweiger}(2006)}]{stollEasySpinComprehensiveSoftware2006}%
  \BibitemOpen
  \bibfield  {author} {\bibinfo {author} {\bibfnamefont {S.}~\bibnamefont {Stoll}}\ and\ \bibinfo {author} {\bibfnamefont {A.}~\bibnamefont {Schweiger}},\ }\bibfield  {title} {\bibinfo {title} {{{EasySpin}}, a comprehensive software package for spectral simulation and analysis in {{EPR}}},\ }\href {https://doi.org/10.1016/j.jmr.2005.08.013} {\bibfield  {journal} {\bibinfo  {journal} {Journal of Magnetic Resonance}\ }\textbf {\bibinfo {volume} {178}},\ \bibinfo {pages} {42} (\bibinfo {year} {2006})}\BibitemShut {NoStop}%
\bibitem [{\citenamefont {Nevzorov}\ \emph {et~al.}(2018)\citenamefont {Nevzorov}, \citenamefont {Milikisiyants}, \citenamefont {Marek},\ and\ \citenamefont {Smirnov}}]{nevzorovMultiresonantPhotonicBandgap2018}%
  \BibitemOpen
  \bibfield  {author} {\bibinfo {author} {\bibfnamefont {A.~A.}\ \bibnamefont {Nevzorov}}, \bibinfo {author} {\bibfnamefont {S.}~\bibnamefont {Milikisiyants}}, \bibinfo {author} {\bibfnamefont {A.~N.}\ \bibnamefont {Marek}},\ and\ \bibinfo {author} {\bibfnamefont {A.~I.}\ \bibnamefont {Smirnov}},\ }\bibfield  {title} {\bibinfo {title} {Multi-resonant photonic band-gap/saddle coil {{DNP}} probehead for static solid state {{NMR}} of microliter volume samples},\ }\href {https://doi.org/10.1016/j.jmr.2018.10.010} {\bibfield  {journal} {\bibinfo  {journal} {Journal of Magnetic Resonance}\ }\textbf {\bibinfo {volume} {297}},\ \bibinfo {pages} {113} (\bibinfo {year} {2018})}\BibitemShut {NoStop}%
\bibitem [{\citenamefont {Palani}\ \emph {et~al.}(2024)\citenamefont {Palani}, \citenamefont {Mardini}, \citenamefont {Quan}, \citenamefont {Ouyang}, \citenamefont {Mishra},\ and\ \citenamefont {Griffin}}]{palaniDynamicNuclearPolarization2024}%
  \BibitemOpen
  \bibfield  {author} {\bibinfo {author} {\bibfnamefont {R.~S.}\ \bibnamefont {Palani}}, \bibinfo {author} {\bibfnamefont {M.}~\bibnamefont {Mardini}}, \bibinfo {author} {\bibfnamefont {Y.}~\bibnamefont {Quan}}, \bibinfo {author} {\bibfnamefont {Y.}~\bibnamefont {Ouyang}}, \bibinfo {author} {\bibfnamefont {A.}~\bibnamefont {Mishra}},\ and\ \bibinfo {author} {\bibfnamefont {R.~G.}\ \bibnamefont {Griffin}},\ }\bibfield  {title} {\bibinfo {title} {Dynamic {{Nuclear Polarization}} with {{P1 Centers}} in {{Diamond}}},\ }\href {https://doi.org/10.1021/acs.jpclett.4c02612} {\bibfield  {journal} {\bibinfo  {journal} {J. Phys. Chem. Lett.}\ ,\ \bibinfo {pages} {11504}} (\bibinfo {year} {2024})}\BibitemShut {NoStop}%
\bibitem [{\citenamefont {Equbal}\ \emph {et~al.}(2020)\citenamefont {Equbal}, \citenamefont {Tagami},\ and\ \citenamefont {Han}}]{equbalBalancingDipolarExchange2020}%
  \BibitemOpen
  \bibfield  {author} {\bibinfo {author} {\bibfnamefont {A.}~\bibnamefont {Equbal}}, \bibinfo {author} {\bibfnamefont {K.}~\bibnamefont {Tagami}},\ and\ \bibinfo {author} {\bibfnamefont {S.}~\bibnamefont {Han}},\ }\bibfield  {title} {\bibinfo {title} {Balancing dipolar and exchange coupling in biradicals to maximize cross effect dynamic nuclear polarization},\ }\href {https://doi.org/10.1039/D0CP02051F} {\bibfield  {journal} {\bibinfo  {journal} {Phys. Chem. Chem. Phys.}\ }\textbf {\bibinfo {volume} {22}},\ \bibinfo {pages} {13569} (\bibinfo {year} {2020})}\BibitemShut {NoStop}%
\bibitem [{\citenamefont {Equbal}\ \emph {et~al.}(2018)\citenamefont {Equbal}, \citenamefont {Li}, \citenamefont {Leavesley}, \citenamefont {Huang}, \citenamefont {Rajca}, \citenamefont {Rajca},\ and\ \citenamefont {Han}}]{equbalTruncatedCrossEffect2018}%
  \BibitemOpen
  \bibfield  {author} {\bibinfo {author} {\bibfnamefont {A.}~\bibnamefont {Equbal}}, \bibinfo {author} {\bibfnamefont {Y.}~\bibnamefont {Li}}, \bibinfo {author} {\bibfnamefont {A.}~\bibnamefont {Leavesley}}, \bibinfo {author} {\bibfnamefont {S.}~\bibnamefont {Huang}}, \bibinfo {author} {\bibfnamefont {S.}~\bibnamefont {Rajca}}, \bibinfo {author} {\bibfnamefont {A.}~\bibnamefont {Rajca}},\ and\ \bibinfo {author} {\bibfnamefont {S.}~\bibnamefont {Han}},\ }\bibfield  {title} {\bibinfo {title} {Truncated {{Cross Effect Dynamic Nuclear Polarization}}: {{An Overhauser Effect Doppelg\"anger}}},\ }\href {https://doi.org/10.1021/acs.jpclett.8b00751} {\bibfield  {journal} {\bibinfo  {journal} {J. Phys. Chem. Lett.}\ }\textbf {\bibinfo {volume} {9}},\ \bibinfo {pages} {2175} (\bibinfo {year} {2018})}\BibitemShut {NoStop}%
\bibitem [{\citenamefont {Kisselev}(1995)}]{kisselevModulationEffectDynamic1995}%
  \BibitemOpen
  \bibfield  {author} {\bibinfo {author} {\bibfnamefont {Y.}~\bibnamefont {Kisselev}},\ }\bibfield  {title} {\bibinfo {title} {The modulation effect on the dynamic polarization of nuclear spins},\ }\href {https://doi.org/10.1016/0168-9002(94)01453-1} {\bibfield  {journal} {\bibinfo  {journal} {Nuclear Instruments and Methods in Physics Research Section A: Accelerators, Spectrometers, Detectors and Associated Equipment}\ }\textbf {\bibinfo {volume} {356}},\ \bibinfo {pages} {99} (\bibinfo {year} {1995})}\BibitemShut {NoStop}%
\bibitem [{\citenamefont {Thurber}\ \emph {et~al.}(2010)\citenamefont {Thurber}, \citenamefont {Yau},\ and\ \citenamefont {Tycko}}]{thurberLowtemperatureDynamicNuclear2010}%
  \BibitemOpen
  \bibfield  {author} {\bibinfo {author} {\bibfnamefont {K.~R.}\ \bibnamefont {Thurber}}, \bibinfo {author} {\bibfnamefont {W.-M.}\ \bibnamefont {Yau}},\ and\ \bibinfo {author} {\bibfnamefont {R.}~\bibnamefont {Tycko}},\ }\bibfield  {title} {\bibinfo {title} {Low-temperature dynamic nuclear polarization at 9.{{4T}} with a {{30mW}} microwave source},\ }\href {https://doi.org/10.1016/j.jmr.2010.03.016} {\bibfield  {journal} {\bibinfo  {journal} {Journal of Magnetic Resonance}\ }\textbf {\bibinfo {volume} {204}},\ \bibinfo {pages} {303} (\bibinfo {year} {2010})}\BibitemShut {NoStop}%
\bibitem [{\citenamefont {Bornet}\ \emph {et~al.}(2014)\citenamefont {Bornet}, \citenamefont {Milani}, \citenamefont {Vuichoud}, \citenamefont {Perez~Linde}, \citenamefont {Bodenhausen},\ and\ \citenamefont {Jannin}}]{bornetMicrowaveFrequencyModulation2014}%
  \BibitemOpen
  \bibfield  {author} {\bibinfo {author} {\bibfnamefont {A.}~\bibnamefont {Bornet}}, \bibinfo {author} {\bibfnamefont {J.}~\bibnamefont {Milani}}, \bibinfo {author} {\bibfnamefont {B.}~\bibnamefont {Vuichoud}}, \bibinfo {author} {\bibfnamefont {A.~J.}\ \bibnamefont {Perez~Linde}}, \bibinfo {author} {\bibfnamefont {G.}~\bibnamefont {Bodenhausen}},\ and\ \bibinfo {author} {\bibfnamefont {S.}~\bibnamefont {Jannin}},\ }\bibfield  {title} {\bibinfo {title} {Microwave frequency modulation to enhance {{Dissolution Dynamic Nuclear Polarization}}},\ }\href {https://doi.org/10.1016/j.cplett.2014.04.013} {\bibfield  {journal} {\bibinfo  {journal} {Chemical Physics Letters}\ }\textbf {\bibinfo {volume} {602}},\ \bibinfo {pages} {63} (\bibinfo {year} {2014})}\BibitemShut {NoStop}%
\bibitem [{\citenamefont {Terblanche}\ and\ \citenamefont {Reynhardt}(2000)}]{terblancheRoomtemperatureFieldDependence2000}%
  \BibitemOpen
  \bibfield  {author} {\bibinfo {author} {\bibfnamefont {C.~J.}\ \bibnamefont {Terblanche}}\ and\ \bibinfo {author} {\bibfnamefont {E.~C.}\ \bibnamefont {Reynhardt}},\ }\bibfield  {title} {\bibinfo {title} {Room-temperature field dependence of the electron spin--lattice relaxation times of paramagnetic {{P1}} and {{P2}} centers in diamond},\ }\href {https://doi.org/10.1016/S0009-2614(00)00411-5} {\bibfield  {journal} {\bibinfo  {journal} {Chemical Physics Letters}\ }\textbf {\bibinfo {volume} {322}},\ \bibinfo {pages} {273} (\bibinfo {year} {2000})}\BibitemShut {NoStop}%
\bibitem [{\citenamefont {Reynhardt}\ \emph {et~al.}(1998)\citenamefont {Reynhardt}, \citenamefont {High},\ and\ \citenamefont {Van~Wyk}}]{reynhardtTemperatureDependenceSpinspin1998}%
  \BibitemOpen
  \bibfield  {author} {\bibinfo {author} {\bibfnamefont {E.~C.}\ \bibnamefont {Reynhardt}}, \bibinfo {author} {\bibfnamefont {G.~L.}\ \bibnamefont {High}},\ and\ \bibinfo {author} {\bibfnamefont {J.~A.}\ \bibnamefont {Van~Wyk}},\ }\bibfield  {title} {\bibinfo {title} {Temperature dependence of spin-spin and spin-lattice relaxation times of paramagnetic nitrogen defects in diamond},\ }\href {https://doi.org/10.1063/1.477511} {\bibfield  {journal} {\bibinfo  {journal} {The Journal of Chemical Physics}\ }\textbf {\bibinfo {volume} {109}},\ \bibinfo {pages} {8471} (\bibinfo {year} {1998})}\BibitemShut {NoStop}%
\bibitem [{\citenamefont {Henstra}\ \emph {et~al.}(1988)\citenamefont {Henstra}, \citenamefont {Dirksen},\ and\ \citenamefont {Wenckebach}}]{henstraEnhancedDynamicNuclear1988}%
  \BibitemOpen
  \bibfield  {author} {\bibinfo {author} {\bibfnamefont {A.}~\bibnamefont {Henstra}}, \bibinfo {author} {\bibfnamefont {P.}~\bibnamefont {Dirksen}},\ and\ \bibinfo {author} {\bibfnamefont {{\relax W.Th}.}~\bibnamefont {Wenckebach}},\ }\bibfield  {title} {\bibinfo {title} {Enhanced dynamic nuclear polarization by the integrated solid effect},\ }\href {https://doi.org/10.1016/0375-9601(88)90950-4} {\bibfield  {journal} {\bibinfo  {journal} {Physics Letters A}\ }\textbf {\bibinfo {volume} {134}},\ \bibinfo {pages} {134} (\bibinfo {year} {1988})}\BibitemShut {NoStop}%
\bibitem [{\citenamefont {Henstra}\ and\ \citenamefont {Wenckebach}(2014)}]{henstraDynamicNuclearPolarisation2014}%
  \BibitemOpen
  \bibfield  {author} {\bibinfo {author} {\bibfnamefont {A.}~\bibnamefont {Henstra}}\ and\ \bibinfo {author} {\bibfnamefont {{\relax W.Th}.}~\bibnamefont {Wenckebach}},\ }\bibfield  {title} {\bibinfo {title} {Dynamic nuclear polarisation via the integrated solid effect {{I}}: Theory},\ }\href {https://doi.org/10.1080/00268976.2013.861936} {\bibfield  {journal} {\bibinfo  {journal} {Molecular Physics}\ }\textbf {\bibinfo {volume} {112}},\ \bibinfo {pages} {1761} (\bibinfo {year} {2014})}\BibitemShut {NoStop}%
\bibitem [{\citenamefont {Can}\ \emph {et~al.}(2018)\citenamefont {Can}, \citenamefont {McKay}, \citenamefont {Weber}, \citenamefont {Yang}, \citenamefont {Dubroca}, \citenamefont {Van~Tol}, \citenamefont {Hill},\ and\ \citenamefont {Griffin}}]{canFrequencySweptIntegratedStretched2018}%
  \BibitemOpen
  \bibfield  {author} {\bibinfo {author} {\bibfnamefont {T.~V.}\ \bibnamefont {Can}}, \bibinfo {author} {\bibfnamefont {J.~E.}\ \bibnamefont {McKay}}, \bibinfo {author} {\bibfnamefont {R.~T.}\ \bibnamefont {Weber}}, \bibinfo {author} {\bibfnamefont {C.}~\bibnamefont {Yang}}, \bibinfo {author} {\bibfnamefont {T.}~\bibnamefont {Dubroca}}, \bibinfo {author} {\bibfnamefont {J.}~\bibnamefont {Van~Tol}}, \bibinfo {author} {\bibfnamefont {S.}~\bibnamefont {Hill}},\ and\ \bibinfo {author} {\bibfnamefont {R.~G.}\ \bibnamefont {Griffin}},\ }\bibfield  {title} {\bibinfo {title} {Frequency-{{Swept Integrated}} and {{Stretched Solid Effect Dynamic Nuclear Polarization}}},\ }\href {https://doi.org/10.1021/acs.jpclett.8b01002} {\bibfield  {journal} {\bibinfo  {journal} {J. Phys. Chem. Lett.}\ }\textbf {\bibinfo {volume} {9}},\ \bibinfo {pages} {3187} (\bibinfo {year} {2018})}\BibitemShut {NoStop}%
\bibitem [{\citenamefont {Kundu}\ \emph {et~al.}(2019)\citenamefont {Kundu}, \citenamefont {{Mentink-Vigier}}, \citenamefont {Feintuch},\ and\ \citenamefont {Vega}}]{kunduDNPMechanisms2019}%
  \BibitemOpen
  \bibfield  {author} {\bibinfo {author} {\bibfnamefont {K.}~\bibnamefont {Kundu}}, \bibinfo {author} {\bibfnamefont {F.}~\bibnamefont {{Mentink-Vigier}}}, \bibinfo {author} {\bibfnamefont {A.}~\bibnamefont {Feintuch}},\ and\ \bibinfo {author} {\bibfnamefont {S.}~\bibnamefont {Vega}},\ }\bibfield  {title} {\bibinfo {title} {{{DNP}} mechanisms},\ }\href {https://doi.org/10.1002/9780470034590.emrstm1550} {\bibfield  {journal} {\bibinfo  {journal} {eMagRes}\ }\textbf {\bibinfo {volume} {8}},\ \bibinfo {pages} {295} (\bibinfo {year} {2019})}\BibitemShut {NoStop}%
\bibitem [{\citenamefont {Leavesley}\ \emph {et~al.}(2017)\citenamefont {Leavesley}, \citenamefont {Shimon}, \citenamefont {Siaw}, \citenamefont {Feintuch}, \citenamefont {Goldfarb}, \citenamefont {Vega}, \citenamefont {Kaminker},\ and\ \citenamefont {Han}}]{leavesleyEffectElectronSpectral2017}%
  \BibitemOpen
  \bibfield  {author} {\bibinfo {author} {\bibfnamefont {A.}~\bibnamefont {Leavesley}}, \bibinfo {author} {\bibfnamefont {D.}~\bibnamefont {Shimon}}, \bibinfo {author} {\bibfnamefont {T.~A.}\ \bibnamefont {Siaw}}, \bibinfo {author} {\bibfnamefont {A.}~\bibnamefont {Feintuch}}, \bibinfo {author} {\bibfnamefont {D.}~\bibnamefont {Goldfarb}}, \bibinfo {author} {\bibfnamefont {S.}~\bibnamefont {Vega}}, \bibinfo {author} {\bibfnamefont {I.}~\bibnamefont {Kaminker}},\ and\ \bibinfo {author} {\bibfnamefont {S.}~\bibnamefont {Han}},\ }\bibfield  {title} {\bibinfo {title} {Effect of electron spectral diffusion on static dynamic nuclear polarization at 7 {{Tesla}}},\ }\href {https://doi.org/10.1039/C6CP06893F} {\bibfield  {journal} {\bibinfo  {journal} {Phys. Chem. Chem. Phys.}\ }\textbf {\bibinfo {volume} {19}},\ \bibinfo {pages} {3596} (\bibinfo {year} {2017})}\BibitemShut {NoStop}%
\bibitem [{\citenamefont {Shimon}\ and\ \citenamefont {Kaminker}(2022)}]{shimonTransitionSolidEffect2022}%
  \BibitemOpen
  \bibfield  {author} {\bibinfo {author} {\bibfnamefont {D.}~\bibnamefont {Shimon}}\ and\ \bibinfo {author} {\bibfnamefont {I.}~\bibnamefont {Kaminker}},\ }\bibfield  {title} {\bibinfo {title} {A transition from solid effect to indirect cross effect with broadband microwave irradiation},\ }\href {https://doi.org/10.1039/D1CP05096F} {\bibfield  {journal} {\bibinfo  {journal} {Phys. Chem. Chem. Phys.}\ }\textbf {\bibinfo {volume} {24}},\ \bibinfo {pages} {7311} (\bibinfo {year} {2022})}\BibitemShut {NoStop}%
\bibitem [{\citenamefont {Abragam}\ and\ \citenamefont {Bleaney}(1970)}]{abragamElectronParamagneticResonance1970}%
  \BibitemOpen
  \bibfield  {author} {\bibinfo {author} {\bibfnamefont {A.}~\bibnamefont {Abragam}}\ and\ \bibinfo {author} {\bibfnamefont {B.}~\bibnamefont {Bleaney}},\ }\href@noop {} {\emph {\bibinfo {title} {Electron {{Paramagnetic Resonance}} of {{Transition Ions}}}}}\ (\bibinfo  {publisher} {Oxford University Press},\ \bibinfo {year} {1970})\BibitemShut {NoStop}%
\bibitem [{\citenamefont {Hovav}\ \emph {et~al.}(2015)\citenamefont {Hovav}, \citenamefont {Shimon}, \citenamefont {Kaminker}, \citenamefont {Feintuch}, \citenamefont {Goldfarb},\ and\ \citenamefont {Vega}}]{hovavEffectsElectronPolarization2015}%
  \BibitemOpen
  \bibfield  {author} {\bibinfo {author} {\bibfnamefont {Y.}~\bibnamefont {Hovav}}, \bibinfo {author} {\bibfnamefont {D.}~\bibnamefont {Shimon}}, \bibinfo {author} {\bibfnamefont {I.}~\bibnamefont {Kaminker}}, \bibinfo {author} {\bibfnamefont {A.}~\bibnamefont {Feintuch}}, \bibinfo {author} {\bibfnamefont {D.}~\bibnamefont {Goldfarb}},\ and\ \bibinfo {author} {\bibfnamefont {S.}~\bibnamefont {Vega}},\ }\bibfield  {title} {\bibinfo {title} {Effects of the electron polarization on dynamic nuclear polarization in solids},\ }\href {https://doi.org/10.1039/C4CP05625F} {\bibfield  {journal} {\bibinfo  {journal} {Phys. Chem. Chem. Phys.}\ }\textbf {\bibinfo {volume} {17}},\ \bibinfo {pages} {6053} (\bibinfo {year} {2015})}\BibitemShut {NoStop}%
\bibitem [{\citenamefont {Hovav}\ \emph {et~al.}(2014)\citenamefont {Hovav}, \citenamefont {Feintuch}, \citenamefont {Vega},\ and\ \citenamefont {Goldfarb}}]{hovavDynamicNuclearPolarization2014}%
  \BibitemOpen
  \bibfield  {author} {\bibinfo {author} {\bibfnamefont {Y.}~\bibnamefont {Hovav}}, \bibinfo {author} {\bibfnamefont {A.}~\bibnamefont {Feintuch}}, \bibinfo {author} {\bibfnamefont {S.}~\bibnamefont {Vega}},\ and\ \bibinfo {author} {\bibfnamefont {D.}~\bibnamefont {Goldfarb}},\ }\bibfield  {title} {\bibinfo {title} {Dynamic nuclear polarization using frequency modulation at 3.34 {{T}}},\ }\href {https://doi.org/10.1016/j.jmr.2013.10.025} {\bibfield  {journal} {\bibinfo  {journal} {Journal of Magnetic Resonance}\ }\textbf {\bibinfo {volume} {238}},\ \bibinfo {pages} {94} (\bibinfo {year} {2014})}\BibitemShut {NoStop}%
\bibitem [{\citenamefont {Guy}\ \emph {et~al.}(2015)\citenamefont {Guy}, \citenamefont {Zhu},\ and\ \citenamefont {Ramanathan}}]{guyDesignCharacterizationWband2015}%
  \BibitemOpen
  \bibfield  {author} {\bibinfo {author} {\bibfnamefont {M.~L.}\ \bibnamefont {Guy}}, \bibinfo {author} {\bibfnamefont {L.}~\bibnamefont {Zhu}},\ and\ \bibinfo {author} {\bibfnamefont {C.}~\bibnamefont {Ramanathan}},\ }\bibfield  {title} {\bibinfo {title} {Design and characterization of a {{W-band}} system for modulated {{DNP}} experiments},\ }\href {https://doi.org/10.1016/j.jmr.2015.09.011} {\bibfield  {journal} {\bibinfo  {journal} {Journal of Magnetic Resonance}\ }\textbf {\bibinfo {volume} {261}},\ \bibinfo {pages} {11} (\bibinfo {year} {2015})}\BibitemShut {NoStop}%
\bibitem [{\citenamefont {Siaw}\ \emph {et~al.}(2016)\citenamefont {Siaw}, \citenamefont {Leavesley}, \citenamefont {Lund}, \citenamefont {Kaminker},\ and\ \citenamefont {Han}}]{siawVersatileModularQuasi2016}%
  \BibitemOpen
  \bibfield  {author} {\bibinfo {author} {\bibfnamefont {T.~A.}\ \bibnamefont {Siaw}}, \bibinfo {author} {\bibfnamefont {A.}~\bibnamefont {Leavesley}}, \bibinfo {author} {\bibfnamefont {A.}~\bibnamefont {Lund}}, \bibinfo {author} {\bibfnamefont {I.}~\bibnamefont {Kaminker}},\ and\ \bibinfo {author} {\bibfnamefont {S.}~\bibnamefont {Han}},\ }\bibfield  {title} {\bibinfo {title} {A versatile and modular quasi optics-based 200 {{GHz}} dual dynamic nuclear polarization and electron paramagnetic resonance instrument},\ }\href {https://doi.org/10.1016/j.jmr.2015.12.012} {\bibfield  {journal} {\bibinfo  {journal} {Journal of Magnetic Resonance}\ }\textbf {\bibinfo {volume} {264}},\ \bibinfo {pages} {131} (\bibinfo {year} {2016})}\BibitemShut {NoStop}%
\end{thebibliography}%

\end{document}